\documentclass[journal, twoside, final]{IEEEtran}

\usepackage{cite}
\usepackage{graphicx}
\usepackage{amssymb, amsmath, amsthm}
\usepackage{amsfonts}
\usepackage{multirow}
\usepackage{mathrsfs}
\usepackage{color}
\usepackage{url}
\usepackage{flushend}
\usepackage{setspace}
\usepackage{placeins}
\usepackage{booktabs, multicol, multirow}
\usepackage{acronym}
\usepackage{bm}
\usepackage{placeins}
\usepackage{extarrows}
\usepackage{color}
\allowdisplaybreaks

\newtheorem{theorem}{Theorem}
\newtheorem{remark}{Remark}
\newtheorem{corollary}{Corollary}

\newmuskip\pFqmuskip
\newcommand*\pFq[6][8]{%
	\begingroup 
	\pFqmuskip=#1mu\relax
	\mathcode`\,=\string"8000
	\begingroup\lccode`\~=`\,
	\lowercase{\endgroup\let~}\pFqcomma
	{}_{#2}F_{#3}{\left[\genfrac..{0pt}{}{#4}{#5};#6\right]}%
	\endgroup
}
\newcommand{\pFqcomma}{\mskip\pFqmuskip}
\newcommand{\RNum}[1]{\uppercase\expandafter{\romannumeral #1\relax}}

\begin{document}

\title{Coordinated Multi-Point Transmission: A Poisson-Delaunay Triangulation Based Approach}
\author{Yan Li, Minghua~Xia,~\IEEEmembership{Member,~IEEE}, and Sonia~A\"{i}ssa,~\IEEEmembership{Fellow,~IEEE}
\thanks{Manuscript received April 10, 2019; revised September 16, 2019 and December 24, 2019; accepted January 16, 2020. This work was supported in part by the National Natural Science Foundation of China under Grant 61671488,  in part by the Major Science and Technology Special Project of Guangdong Province under Grant 2018B010114001, in part by the Fundamental Research Funds for the Central Universities under Grant 191gjc04, and in part by a Discovery Grant from the Natural Sciences and Engineering Research Council (NSERC) of Canada. The associate editor coordinating the review of this paper and approving it for publication was A. Zaidi.}
\thanks{Y. Li and M. Xia are with the School of Electronics and Information Technology, Sun Yat-sen University, Guangzhou, 510006, China. M. Xia is also with Southern Marine Science and Engineering Guangdong Laboratory (Zhuhai) (e-mail: liyan228@mail2.sysu.edu.cn, xiamingh@mail.sysu.edu.cn).}
\thanks{S. A\"{i}ssa is with the Institut National de la Recherche Scientifique (INRS-EMT), University of Quebec, Montreal, QC, H5A 1K6, Canada (e-mail: aissa@emt.inrs.ca).}
\thanks{Color versions of one or more of the figures in this paper are available online at http://ieeexplore.ieee.org.}
\thanks{Digital Object Identifier }
}
	
\markboth{IEEE Transactions on Wireless Communications} {Xia \MakeLowercase{\textit{et al.}}: Coordinated Multi-Point Transmission: A Poisson-Delaunay Triangulation Based Approach}

\maketitle

\IEEEpubid{\begin{minipage}{\textwidth} \ \\[12pt] \centering 1536-1276 \copyright\ 2018 IEEE. Translations and content mining are permitted for academic research only. Personal use is also permitted, \\
but republication/redistribution requires IEEE permission. See \url{http://www.ieee.org/publications_standards/publications/rights/index.html for more information}.\end{minipage}}

\IEEEpubidadjcol
	
\begin{abstract}
\noindent Coordinated multi-point (CoMP) transmission is a cooperating technique among base stations (BSs) in a cellular network, with outstanding capability at inter-cell interference (ICI) mitigation. ICI is a dominant source of error, and has detrimental effects on system performance if not managed properly. Based on the theory of Poisson-Delaunay triangulation, this paper proposes a novel analytical model for CoMP operation in cellular networks. Unlike the conventional CoMP operation that is dynamic and needs on-line updating occasionally, the proposed approach enables the cooperating BS set of a user equipment (UE) to be fixed and off-line determined according to the location information of BSs. By using the theory of stochastic geometry, the coverage probability and spectral efficiency of a typical UE are analyzed, and simulation results corroborate the effectiveness of the proposed CoMP scheme and the developed performance analysis.
\end{abstract}

\section{Introduction}
\label{Section-Introduction}
\IEEEPARstart{C}{oordinated} multi-point (CoMP) transmission and reception is considered for the 3rd Generation Partnership Project (3GPP) long term evolution advanced (LTE-A) as a promising technique to mitigate inter-cell interference, thereby improving the system coverage, the spectral efficiency and in particular the quality-of-service (QoS) of cell-edge user equipments (UEs) in cellular networks. In the state-of-the-art technical report for physical layer aspects of the study item ``Coordinated multi-point operation for LTE'', namely, 3GPP TR 36.819 \cite{3GPPTR36.819}, two major CoMP strategies are highlighted: coordinated scheduling/beamforming (CS/CB) and joint processing. In the CS/CB strategy, data for a UE is only available at one point of the CoMP cooperation set in a time-frequency resource block. In the joint processing strategy, on the other and, data for a UE is available at more than one point in the CoMP cooperating set. Clearly, the joint processing strategy outperforms the CS/CB, but at the cost of higher backhaul load. In practice, the joint processing strategy has two major implementation schemes: joint transmission, and dynamic point selection/muting \cite{3GPPTR36.819}. In the former, multiple points simultaneously transmit data to a UE in a time-frequency resource block so as to improve data throughput and/or decrease outage probability. As for the latter, although data is simultaneously available at multiple points, only one point out of the cooperation set transmits data to a UE. In this paper, both implementation schemes of joint processing will be examined.
	
\IEEEpubidadjcol

In the open literature, there are two distinct methodologies to investigate the performance of CoMP in cellular networks. One is the classic {\it deterministic approach}, which is based on the widely used regular hexagonal cellular model. This method is simple yet highly idealized and, hence, inaccurate in practice. To better reflect the actual deployment of base stations (BSs), the theory of stochastic geometry was in recent years introduced to model and analyze cellular networks, yielding the novel {\it stochastic approach} \cite{Elsawy2017Modeling}, where a Poisson point process (PPP) is used to describe the distribution of BSs while UEs are uniformly distributed in the coverage area of the network. Each UE is then associated with a target BS by using the nearest-neighbor criterion and, accordingly, the polygonal boundaries around BSs form a Poisson-Voronoi tessellation \cite{Chiu13}. Using the theory of Poisson-Voronoi tessellation, when no CoMP transmission is considered among BSs, the coverage probability of a typical UE inside Poisson-Voronoi cells was analyzed in \cite{AndrewsTCOM1111}, and the performance of the worst-case users at the vertices of Poisson-Voronoi cells was investigated in \cite{JungCL1308}. By using CoMP, the performance of the worst-case users at the vertices of Poisson-Voronoi cells was studied in \cite{NigamTCOM14s}. The performance of a dynamic coordinated beamforming was characterized in \cite{6851166}, where each UE is assumed to communicate only with the nearest BS in its CoMP cooperation set. A dynamic interference nulling strategy for small-cell networks was proposed in \cite{7038201}, and its average data rate was analyzed in \cite{8314102}. More recently, the stochastic approach was also applied to study heterogeneous cellular networks. For instance, it was validated that the spatial distribution of macro- and micro-cell BSs can be modeled as the superposition of two independent PPPs \cite{Wu2014Spatial}. Further, concerning CoMP among BSs, the coverage probabilities for a typical UE in heterogeneous downlink networks was studied in \cite{Nigam2015Spatiotemporal, 8100895}, and the signal-to-interference ratio (SIR) meta distribution for both the general UE and the worst-case UE under the Poisson multiple-tier cellular networks was analyzed in \cite{8115171}. Most recently, stochastic geometry was integrated with optimization theory for optimal design and performance analysis of cellular networks, see e.g., \cite{Chen2016Area, Li2016Optimization, Chang2017Energy, Shojaeifard2017Stochastic}.

When the theory of stochastic geometry is applied to cellular networks, the coverage area of a network is usually tessellated by Poisson-Voronoi tessellation, where one UE is associated with its nearest BS. However, since a typical cell of Poisson-Voronoi tessellation is an irregular polygon with the number of edges varying from $3$ to $13$, some basic features of a typical cell, for instance, the probability density function (PDF) of its area, is still unknown so far. This hinders the analytical performance evaluation of cellular networks. As the dual diagram of Poisson-Voronoi tessellation, in contrast, Poisson-Delaunay triangulation has regular triangular cells, namely, a typical cell of Poisson-Delaunay triangulation is always triangular. This regularity makes the applications of Poisson-Delaunay triangulation more mathematically tractable \cite{Triangulations10}. In our recent work \cite{Xia2018Un}, Poisson-Delaunay triangulation was used to model cellular networks and a novel CoMP transmission scheme was proposed. Unlike the conventional user-centric CoMP operations, such as \cite{AndrewsTCOM1111, JungCL1308, NigamTCOM14s, 6851166, 7038201, 8314102, 8100895} where on-line searching and feedback overhead are necessary to determine the cooperation set of a UE, one of the key features of the said CoMP scheme \cite{Xia2018Un} is that the set of cooperative BSs pertaining to any UE is fixed and can be off-line determined once the geographic locations of BSs are known, which is feasible in real-world cellular networks. As a companion work to \cite{Xia2018Un}, this paper investigates the network performance of Poisson-Delaunay triangulation based CoMP transmission, in terms of the coverage probability and the spectral efficiency.

Specifically, this paper studies the performance of CoMP transmission based on Poisson-Delaunay triangulation. Since the UEs at the vertices of conventional Poisson-Voronoi tessellation suffer the worst QoS, to characterize this QoS, a typical UE is intentionally chosen to be located at a vertex of a Poisson-Voronoi cell, which exactly has equal distances from the three neighbouring BSs at the vertices of the dual Poisson-Delaunay triangle. On the other hand, two joint processing schemes, namely, joint transmission and dynamic point selection/muting, are employed at BSs. By using the theory of stochastic geometry, the coverage probability and spectral efficiency of a typical UE are analytically derived. Monte Carlo simulation results are also provided and corroborate the effectiveness of the proposed CoMP scheme and the corresponding performance analysis.

The rest of this paper is organized as follows. Section~\ref{Section-SystemModel} describes the system model and the principle of constructing the cooperation set of a UE. Then, Sections~\ref{Section-JT} and \ref{Section-DPSM} are devoted to JT and dynamic point selection/muting techniques at BSs, respectively, where in each case the coverage probability and spectral efficiency of a typical UE are explicitly derived. Moreover, for comparison purposes, the performance of transmission without CoMP is investigated. Simulation results are presented and discussed in Section~\ref{Section-Simulation}. Finally, Section~\ref{Section-Conclusion} concludes the work.
	
{\it Notation}: The operator $\mathbb{E}(\cdot)$ means mathematical expectation and ${\rm round}(\cdot)$ takes the nearest integer of a real number. The symbols $\|\bm{x}\|_1$, $\|\bm{x}\|$, and $\bm{x}^H$ denote the $\ell_1$-norm, $\ell_2$-norm, and Hermitian transpose of vector $\bm{x}$, respectively. The function $F^{-1}(x)$ represents the inverse function of $F(x)$, and $\delta(n)$ refers to the Dirac delta function, with $\delta(0) = 1$ and $\delta(n) = 0$ if $n\neq 0$. The symbol $ {n \choose m} = \frac{n!}{m! \, (n-m)!}$ refers to binomial coefficient, with $n!$ being the factorial of a positive integer $n$. The Gamma, incomplete Gamma, and regularized incomplete Gamma functions are defined as $\Gamma(a) \triangleq \int_0^\infty{t^{a-1} \,e^{-t}}\,{\rm d}t$, $\Gamma(a, x) \triangleq \int_x^{\infty} t^{a-1}\,e^{-t}\,{\rm d}t$, and $Q(a, x) \triangleq \Gamma(a, x)/\Gamma(a)$, for all $a>0$, respectively. The generalized hypergeometric function is defined as ${\displaystyle \, {}_{p}F_{q}(a_{1}, \cdots ,a_{p}; b_{1}, \cdots, b_{q}; x) \triangleq \sum_{n=0}^{\infty }{\frac {(a_{1})_{n} \cdots (a_{p})_{n}}{(b_{1})_{n} \cdots (b_{q})_{n}}} \, {\frac {x^{n}}{n!}}}$, with $(a)_n = a(a+1)\cdots(a+n-1)$ if $n \ge 1$ and $(a)_n = 1$ if $n = 0$. Notice that these special functions can be readily computed by using built-in functions in regular numerical softwares, such as Matlab and Mathematica.

\section{Network Model}
\label{Section-SystemModel}
\begin{figure}[!t]
	\centering
	\includegraphics [width=3.25in, clip]{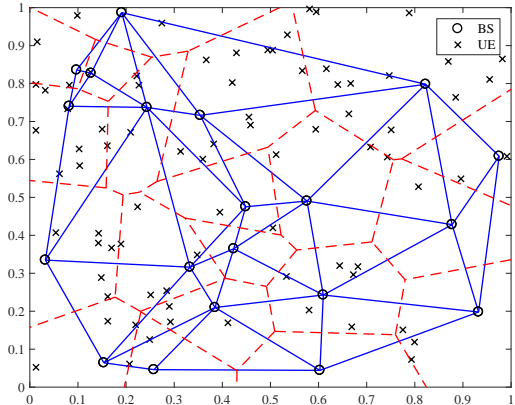}
	\caption{An illustrative cellular network modeled by the Poisson-Voronoi tessellation (polygons with red dash boundaries) or by the dual Poisson-Delaunay triangulation (triangles with blue solid boundaries), with normalized coverage area of one squared kilometers.}
	\label{Fig-1}
\end{figure}

Figure~\ref{Fig-1} illustrates a cellular network where the BSs and the UEs are denoted by the `$\circ$' and `$\times$' marks, respectively. The BSs are assumed to be distributed in a two-dimensional (2D) infinite plane as per a homogeneous PPP, denoted $\Phi$, with intensity $\lambda$. If each UE, uniformly distributed in the plane, is associated to its nearest BS in the sense of Euclidean distance, the resulting polygonal boundaries form a {\it Poisson-Voronoi tessellation}, as shown by the red dash lines in Fig.~\ref{Fig-1}. On the other hand, the dual {\it Poisson-Delaunay triangulation} is illustrated as the triangles with blue solid boundaries. Each red polygon associated with a BS is known as a Poisson-Voronoi cell, while each blue triangle associated with three BSs represents a Poisson-Delaunay cell. Once the locations of BSs are known, the Poisson-Voronoi tessellation and Poisson-Delaunay triangulation are uniquely determined and they are dual Siamese twins \cite{Liebling12}.

To serve UEs in triangular Poisson-Delaunay cells, each BS is assumed to be equipped with a large number of antennas, which enables multiple narrow directional beams as required. At each UE, a single receive antenna is assumed. The analysis in the rest of this paper confines to the single-antenna UE case, but it can be extended to the multi-antenna UE case in a straightforward manner, for example, by treating each UE antenna as a separate UE or using maximum ratio combining at the UE \cite[Section 2.2]{Marzetta16}.

\subsection{Principles to Determine the Cooperation Set of a UE}
\label{Section-CooperatingSet}
Based on the geometric locations of BSs, the Poisson-Delaunay triangulation dual to the Poisson-Voronoi tessellation is uniquely determined and can be efficiently constructed by using, e.g., the radial sweep or divide-and-conquer algorithm \cite[ch. 4]{Hjelle06}. Then, for each UE, the CoMP cooperation set can be readily determined. More specifically,  as shown in Fig.~5a of \cite{Xia2018Un}, if a UE is located inside a Poisson-Delaunay triangular cell, the three BSs at the vertices of the triangle are chosen and form the CoMP cooperating set. On the other hand, if a UE is exactly located on the edge of a triangle as shown in Fig.~5b of \cite{Xia2018Un}, there must be an adjacent triangle which shares the same edge and they both form a quadrilateral (the edge effect of the whole cellular network is ignored due to its large coverage area). Among the four BSs at the vertices of the quadrilateral, the UE on the edge chooses the two BSs at both ends of the edge and a third BS among the remaining two opposite BSs which represents the one closer to the UE, so as to form the CoMP cooperating set.
\subsection{Three Types of UEs}
\label{Subsect_TypesUEs}
\begin{figure}[!t]
	\centering
	\includegraphics [width=3.25in, clip, keepaspectratio]{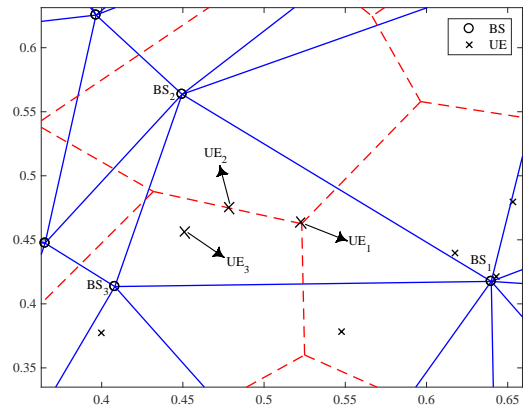}
	\caption{An illustration of three types of UEs, where $\text{UE}_1$ has the same distances from $\text{BS}_1$, $\text{BS}_2$ and $\text{BS}_3$, $\text{UE}_2$ has the same distances from $\text{BS}_1$ and $\text{BS}_2$ but another distance from $\text{BS}_3$, and $\text{UE}_3$ has distinct distances from the three serving BSs.}
	\label{Fig-2}
\end{figure}

According to the Euclidean distances from the three BSs in a CoMP cooperation set determined as per the above principles, all UEs in the network can be classified into three types. Type~\RNum{1} UEs are located at the centroids of triangular cells and each of them is equidistant from its three serving BSs. A Type~\RNum{2} UE is equidistant from two BSs but has another distance from the third BS. Type~\RNum{3} UE has distinct distances from its three serving BSs. For illustration purposes, $\text{UE}_1$, $\text{UE}_2$ and $\text{UE}_3$ in Fig.~\ref{Fig-2} correspond to Type~\RNum{1}, Type~\RNum{2} and Type~\RNum{3}, respectively, with the cooperating BS set consisting of $\text{BS}_1$, $\text{BS}_2$ and $\text{BS}_3$.

Alternatively, by taking a closer look at Fig.~\ref{Fig-2}, it is not hard to recognize that Type~\RNum{1} UEs in the proposed Poisson-Delaunay cells are located at the vertices of the dual Poisson-Voronoi cells, while Type~\RNum{2} UEs are on the edge of Poisson-Voronoi cells and Type~\RNum{3} UEs are inside Poisson-Voronoi cells. Clearly, Types~\RNum{1} and \RNum{2} UEs are indeed the cell-edge users in conventional cellular systems without CoMP operation, which suffer the worst QoS \cite{JungCL1308}. As well-known, by means of CoMP operation, the performance of all UEs can be significantly enhanced. To demonstrate the effectiveness of the Poisson-Delaunay triangulation based CoMP strategy, the rest of this paper focuses on Type~\RNum{1} UEs and analyzes its coverage probability and spectral efficiency. The performance of Types~\RNum{2} and \RNum{3} UEs will be investigated in our future work.

\subsection{Received SIR at a Typical UE}
\label{Section-SINR}
By using the Slivnyak-Mecke theorem \cite[p. 132]{Chiu13}, a typical UE can be assumed to be located at the origin $(0, 0) \in \mathbb{R}^2$ of the 2D plane, without loss of generality.\footnote{The application of Slyvniak's theorem is not straightforward due to the location correlation between a typical UE and its serving BSs. However, for ease of mathematical tractability, we resort to Slyvniak's theorem, and its effect on the accuracy of subsequent analyses are explicitly examined in Section V.}
The three BSs in the CoMP cooperating set, $\Phi_0 = \{\mathrm{A_0, B_0, C_0}\}$, jointly transmit signals to a typical UE whereas the BSs in the $j^\mathrm{th}$ adjacent set, $\Phi_j = \{\mathrm{A_j, B_j, C_j}\}$, for all $j = 1, \cdots, \infty$, are treated as external interfering sources, where $\Phi_0 \cup \Phi_j|_{j=1}^{\infty} = \Phi$. As aforementioned,  each BS in the network is equipped with $M$ transmit antennas while each UE has a single antenna. Consequently, the received signal at a typical UE can be expressed as
\begin{equation} \label{Eq_ReceivedSignal}
	y = \sum_{i \in \Phi_0}{P_i^{\frac{1}{2}} d_i^{-\frac{\alpha}{2}} \bm{h}_i^H \bm{w}_i  x_0}
    + \sum_{j=1}^\infty\sum_{k \in \Phi_j}{P_k^{\frac{1}{2}} d_{k, \, 0}^{-\frac{\alpha}{2}} \bm{h}_{k, \, 0}^{H} \bm{w}_k x_j}
	+ z,
\end{equation}
where $P_i$ denotes the transmit power of the $i^\mathrm{th}$~BS; $d_i$ is the Euclidean distance from the $i^\mathrm{th}$~BS to a typical UE; $\alpha > 2$ is the path-loss exponent; $\bm{h}_i \in \mathbb{C}^{M \times 1}$ stands for the complex channel vector from the $i^\mathrm{th}$ BS to a typical UE and $\bm{w}_i  \in \mathbb{C}^{M \times 1}$ is the precoder used at the $i^\mathrm{th}$ BS; $z$ means the additive white Gaussian noise at a typical UE, with zero mean and variance $\sigma^2$. The parameters $P_k$ and $\bm{w}_k$ in the second term on the right-hand side of \eqref{Eq_ReceivedSignal} denote the transmit power and precoder at the $k^\mathrm{th}$ interfering BS, for all $k \in \Phi_j$, while $d_{k, \, 0}$ and $\bm{h}_{k,\,0}$ refer to the distance and channel vector from the $k^\mathrm{th}$ interfering BS to a typical UE located at the origin, respectively. Further, $x_0$ denotes the desired signal that is jointly transmitted by the BSs in the CoMP cooperation set pertaining to a typical UE, while $x_j$ refers to the interfering signal transmitted by the BSs belonging to the adjacent set $\Phi_j$, for all $j=1, \cdots, \infty$. $x_0$ and $x_j$ are assumed to be independent and identically distributed (i.i.d.) random variables with zero mean and unit variance. Finally, we note that intra-cell interference is not accounted for in \eqref{Eq_ReceivedSignal} since it can be effectively mitigated by using techniques such as orthogonal frequency division multiple access (OFDMA).
	
Since we consider the downlink transmission without power control in a single-tier cellular network, the transmit powers of all BSs are assumed identical and normalized to unity. Also, full downlink channel state information (CSI) is assumed available at BSs interconnected via high-speed optical links. Accordingly, the channel-inverse precoder $\bm{w}_i$ used by the $i^\mathrm{th}$ BS is given by
\begin{equation} \label{Eq_Precoder}
	\bm{w}_i = \frac{\bm{h}_{i}^{}}{\|{\bm{h}_i\|}}.
\end{equation}
Also, as the network performance under study is typically interference-limited, the noise term in \eqref{Eq_ReceivedSignal}, i.e., $z$, is negligible. Thus, by substituting \eqref{Eq_Precoder} into \eqref{Eq_ReceivedSignal}, we can express the instantaneous received signal-to-interference ratio (SIR) at a typical UE as
\begin{equation} \label{Eq_RxSINR}
	\Gamma = \frac{\left|\sum\limits_{i \in \Phi_0}  d_i^{-\frac{\alpha}{2}} \|\bm{h}_i\|\right|^2}
	{\sum\limits_{j=1}^\infty\left|\sum\limits_{k \in \Phi_j} d_{k, \, 0}^{-\frac{\alpha}{2}} \bm{h}_{k, \, 0}^H \frac{\bm{h}_{k}}{\|{\bm{h}_k\|}} \right|^2 }.
\end{equation}	

In the next section, the coverage probability and spectral efficiency of a typical UE in the case of joint trasnmisison among BSs in the cooperation set is investigated, followed by the case of dynamic point selection/muting among BSs.

\begin{figure}[!t]
	 \centering
	\includegraphics [width=3.0in, clip, keepaspectratio]{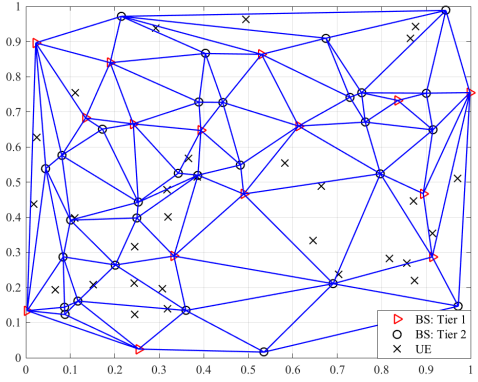}
	\caption{An illustrative application of CoMP transmission based on Poisson-Delaunay triangulation in two-tier heterogeneous networks, where `$\triangleright$' and `$\circ$' denote macro- and micro-cell BSs, respectively, and `$\times$' indicate the UEs.}
	\label{Fig-3}
\end{figure}

\begin{remark}[Extension from single-tier to multi-tier networks]
\label{Remark_1}
	Although a single-tier cellular network is considered in this paper, the idea of CoMP transmission based on Poisson-Delaunay triangulation can be readily applied to multi-tier networks. For instance, Fig.~\ref{Fig-3} shows a two-tier heterogenous cellular network where the macro- and micro-cell BSs are modeled as homogeneous PPPs, denoted $\Phi_1$ and $\Phi_2$ of intensity $\lambda_1$ and $\lambda_2$, respectively. It is well-known that all BSs consisting of macro- and micro-cell BSs can be modeled as the superposition of $\Phi_1$ and $\Phi_2$ \cite{Wu2014Spatial}. With the resulting PPP of intensity $\lambda_1+\lambda_2$, a Poisson-Delaunay triangulation can be obtained, as shown in Fig.~\ref{Fig-3}. Clearly, the BSs at the vertices of a triangle may be either macro- or micro-cell BSs, with which each UE can be associated.
\end{remark}

\section{Joint Transmission}
\label{Section-JT}
In this section, joint transmission (JT) is applied at the three BSs in the cooperation set pertaining to a typical UE, and the coverage probability and spectral efficiency are used to characterize its performance. Mathematically, given an outage threshold on the received SIR at a typical UE, say $\gamma$, the coverage probability is defined as \cite{AndrewsTCOM1111}
\begin{equation} \label{Eq_CoverageProbability}
	\mathcal{P} \triangleq 1 -  \Pr\left\{\Gamma \leq \gamma \right\}.
\end{equation}	
To calculate \eqref{Eq_CoverageProbability}, next we address the distribution characteristics of $\Gamma$ shown in \eqref{Eq_RxSINR}.

\subsection{Received SIR at a Typical UE}
\label{Section-JT-A}
By recalling Fig.~\ref{Fig-3}, a UE at the vertex of a triangular cell, e.g., UE$_1$, is chosen as a typical point and is set to be the origin in $2{\rm D}$ space, which implies that the Euclidean distances between a typical UE and its serving BSs are identical, i.e., $d_i = d$, for all $i \in \Phi_0$. In such a case, \eqref{Eq_RxSINR} reduces to
	\begin{eqnarray}
	\Gamma_\mathrm{1}
	   =     \frac{d^{-\alpha} \left|\sum\limits_{i \in \Phi_0} \|\bm{h}_i\|\right|^2}
	{\sum\limits_{j=1}^\infty\left|\sum\limits_{k \in \Phi_j} d_{k, \, 0}^{-\frac{\alpha}{2}} \bm{h}_{k, \, 0}^H \frac{\bm{h}_{k}}{\|{\bm{h}_k\|}} \right|^2 }
	   =   \frac{d^{-\alpha} \, U}{I_1}, \label{Eq_RxSINR-CaseI}
  \end{eqnarray}
where
\begin{align}
	U  & \triangleq \left|\sum_{i \in \Phi_0} \|\bm{h}_i\|\right|^2  \label{Eq_U},  \\
	I_1  & \triangleq \sum_{j=1}^\infty\left|\sum_{k \in \Phi_j} d_{k, \, 0}^{-\frac{\alpha}{2}} \bm{h}_{k, \, 0}^H \frac{\bm{h}_{k}}{\|{\bm{h}_k\|}} \right|^2. \label{Eq_I}
\end{align}

In the following, the distribution functions of $d$ and $U$ and the Laplace transform of $I_1$ are discussed in sequence.
\subsubsection{The Distribution of the Distance $d$}
Based on the theory of Palm distribution, the PDF of distance $d$ involved in \eqref{Eq_RxSINR-CaseI} is derived in \cite{{Muche2005The}}, and given by
\begin{equation} \label{Eq_PDF_d}
	f_d(x) = 2(\lambda \pi)^2 x^3 \exp\left(-\lambda \pi x^2\right).
\end{equation}
\subsubsection{The Distribution of the Desired Signal $U$}
\label{Section-JT-A-b}
Since each element of the channel vector $\bm{h}_i \in \mathbb{C}^{M \times 1}$ is a complex Gaussian variable with zero mean and unit variance, it is evident that $\|\bm{h}_i\|$ is of Nakagami distribution with PDF given by
\begin{equation}
	f_{\|\bm{h}_i\|}(x) = \frac{2x^{2M-1}}{\Gamma{(M)}  }\exp \left(-x^2\right).
\end{equation}	

Let an intermediate variable $T \triangleq \sum_{i \in \Phi_0} \|\bm{h}_i\| =\sum_{i=1}^{3}\|\bm{h}_i\|$, then $T$ is clearly the sum of three independent Nakagami random variables. In theory, an exact PDF of $T$ is obtainable by using an approach similar to \cite{DharmawansaTCOM0707}, yielding
\begin{align} \label{Eq_PDF-T1}
& f_{T}(x)
	 =  \frac{8\sqrt{\pi} \, \Gamma(2M)}{ \Gamma^{3}(M) \, 2^{4M-1}} \exp\left(-x^2\right) \nonumber\\
	& \quad \times{} \sum\limits_{n = 0}^{\infty}
			\frac{\Gamma(2M+n) \, \Gamma(4M+2n) \, x^{2(3M+n)-1}}{\Gamma(2M+n+\frac{1}{2}) \, \Gamma(6M+2n) \, \Gamma(n+1) \, 2^n}  \nonumber \\
	& \quad   \times{} {_2F_2\left(2M, 4M+2n; 3M+n+\frac{1}{2}, 3M+n; \frac{1}{2}x^2 \right)}.
\end{align}
Albeit accurate, \eqref{Eq_PDF-T1} is too complex to be further processed. For ease of further proceeding, an approximate and accurate PDF of $T$ is used in this paper. Specifically, by using a similar approach to \cite{Filho2004Nakagami}, an approximate PDF of $T$ can be derived and given by
\begin{equation} \label{Eq_PDF_T1-Approx}
	f_{T}(x) \approx \frac{2m^m x^{2m-1}}{\Gamma{(m)}\Omega^m}\exp{\left(-\frac{m x^2}{\Omega}\right)},
\end{equation}
where
\begin{align}
	\Omega & = \mathbb{E}[T^2],  \label{Eq_PDF_T1-Approx-2a} \\
		 m & \triangleq {\rm round}\left( \frac{\Omega^2}{\mathbb{E}[T^4] - \Omega^2} \right).   \label{Eq_PDF_T1-Approx-2b}
\end{align}
To calculate the moments $\mathbb{E}[T^2]$ and $\mathbb{E}[T^4]$ required in \eqref{Eq_PDF_T1-Approx-2a}-\eqref{Eq_PDF_T1-Approx-2b}, by recalling the formula of multinomial expansion, the exact $n^{\rm th}$-order moment of $T$ can be written in terms of the moments of its three components, such that
\begin{align} 
	\mathbb{E}\left[T^n \right]
		&=  \sum_{n_1=0}^{n}\sum_{n_2=0}^{n_1} \binom{n}{n_1}\binom{n_1}{n_2} \mathbb{E}\left[\| \bm{h}_1 \|^{n-n_1} \right] \nonumber\\
		&\quad \times{}        \mathbb{E}\left[\| \bm{h}_2 \|^{n_1-n_2} \right] \mathbb{E}\left[ \| \bm{h}_3 \|^{n_2} \right], \label{Eq-Appro-Moment}
\end{align}
where
\begin{equation} \label{Eq-Appro_Gamma_Moment}
\mathbb{E} \left[\| \bm{h}_i \|^n \right]
	\triangleq \int_{0}^{\infty} x^n  f_{\|\bm{h}_i\|}(x) \, {\rm d}x
	= \frac{\Gamma\left(M+\frac{n}{2}\right)}{\Gamma\left(M \right)}.
\end{equation}

Next, since $U = |T|^2 = T^2$ by noting that $T$ is a non-negative real number, in light of \eqref{Eq_PDF_T1-Approx} the PDF of $U$ can be expressed as
\begin{equation} \label{Eq_PDF_U}
	f_U(x) = \frac{\left(\frac{m}{\Omega} \right)^m}{\Gamma{(m)}}x^{m-1}\exp\left(- \frac{m}{\Omega}x  \right).
\end{equation}
Meanwhile, the complementary cumulative density function (CCDF) of $U$ is readily given by
\begin{align}
	F_U(x) &=  \frac{1}{\Gamma(m)}\Gamma\left(m, \frac{m}{\Omega}x\right) \nonumber \\
		    &=  \sum_{k=0}^{m-1} \frac{1}{k!}\left(\frac{m}{\Omega}x\right)^k \exp\left(-\frac{m}{\Omega}x  \right).  \label{Eq-CCDF_U-2}
\end{align}

\begin{figure}[!t]
	\centering
	\includegraphics[width=3.0in, clip]{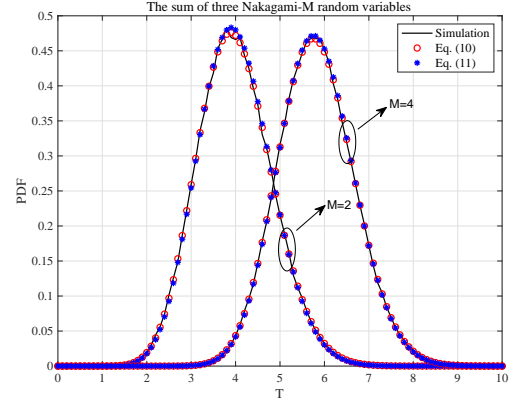}
	\caption{The accuracy of PDFs of $T$ given by Eqs.~\eqref{Eq_PDF-T1} and \eqref{Eq_PDF_T1-Approx}, compared with simulation results.}
	\label{Fig-4}
\end{figure}

\begin{remark}[The accuracy of the approximation given by Eq.~\eqref{Eq_PDF_T1-Approx}]
\label{Remark_2}
For ease of analytical tractability, the value of parameter $m$ is rounded in \eqref{Eq_PDF_T1-Approx-2b} to its nearest integer. For instance, if $M = 2$, after some tedious yet straightforward calculation, we get $m=5.79$. The value $m = 6$ is taken in practice for subsequent numerical calculations such that the finite series expansion shown in \eqref{Eq-CCDF_U-2} holds. Since the value of $m$ is large enough, this approximation yields little deviation from the exact PDF given by \eqref{Eq_PDF-T1}, as illustrated in Fig.~\ref{Fig-4}.
\end{remark}
\subsubsection{The Laplace Transform of the Interference $I_1$}
\label{Subsection_LS_1}

\begin{figure}[!t]
	\centering
	\includegraphics[width=3.25in, clip]{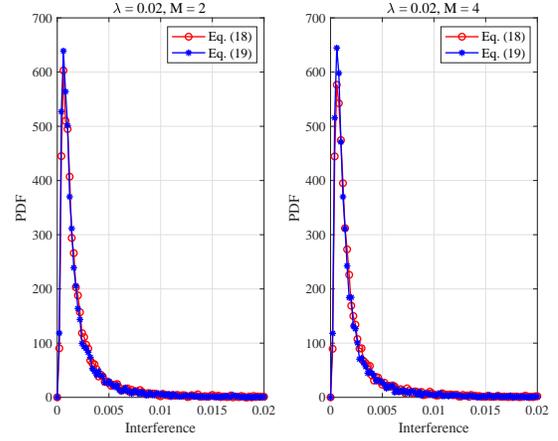}
	\caption{PDFs of the interference calculated as per \eqref{Eq_L_1_a} and \eqref{Eq_L_1_b}, with $\lambda = 0.02$ and $M = 2$ (left) or $M=4$ (right).}
	\label{Fig-5}
\end{figure}

According to \eqref{Eq_I}, the aggregate interference received at a typical UE is given by
\begin{align}
I_1
& \triangleq \sum\limits_{j=1}^\infty\left|\sum\limits_{k \in \Phi_j} d_{k, \, 0}^{-\frac{\alpha}{2}} \bm{h}_{k, \, 0}^H \, \frac{\bm{h}_{k}}{\|{\bm{h}_k\|}} \right|^2  \label{Eq_L_1_a}\\
&\approx \sum\limits_{j =1}^\infty d_{j, \, 0}^{-\alpha}  \left| \sum\limits_{k \in \Phi_j}  \bm{h}_{k, \, 0}^H \, \frac{\bm{h}_{k}}{\|{\bm{h}_k\|}} \right|^2, \label{Eq_L_1_b}
\end{align}
where, for ease of mathematical tractability, the distances from the three BSs in the interfering set $\Phi_j$ to a typical UE are assumed identical and given by $d_{j,\,0}$ in \eqref{Eq_L_1_b}. Intuitively speaking, this assumption is feasible in practice since all the BSs are supposed to be distributed in the infinite 2D plane and, as such, the three BSs in a CoMP cooperation set are relatively close to each other. For illustration purposes, the PDFs of the interference calculated as per \eqref{Eq_L_1_a} and \eqref{Eq_L_1_b} are plotted in Fig.~\ref{Fig-5}, where $\lambda = 0.02$ and $M = 2$ (left-panel) or $M = 4$ (right-panel). As observed, the PDFs of \eqref{Eq_L_1_a} and \eqref{Eq_L_1_b} coincide with each other. Also, Fig.~\ref{Fig-5} shows that the PDF of the interference is surprisingly independent of $M$, i.e., the number of transmit antennas at BSs, as mathematically justified next.

Let $\hat{h}_k \triangleq \sum_{k\in \Phi_j} \bm{h}_{k, \, 0}^H \, \bm{h}_{k}/\|{\bm{h}_k\|}$, since $\bm{h}_{k, \, 0}$ is the channel vector from the $k^\mathrm{th}$ BS to a typical UE whereas $\bm{h}_k$ is the precoder for a local UE, $\bm{h}_{k, \, 0}$ and $\bm{h}_{k}/\|{\bm{h}_k\|}$ are not matched. By noting that $\bm{h}_{k}/\|{\bm{h}_k\|}$ is isotropic, it is clear that $\bm{h}_{k, \, 0}^H \bm{h}_{k}/\|{\bm{h}_k\|}$ is of normalized complex Gaussian distribution with unit mean and, hence, $| \hat{h}_k |^2$ is of exponential distribution with mean $\mu = 3$, independent of the number of transmit antennas at BSs. As a result, substituting $| \hat{h}_k |^2$ into \eqref{Eq_L_1_b}, the Laplace transform of the interference $I_1$ is given by
\begin{align}
L_{I_1}(s)
	& \approx \mathbb{E}_{\hat{\Phi}, \hat{h}}\left[ \exp\left(- s \sum_{k \in \hat{\Phi}} d_{k, \, 0}^{-\alpha} |\hat{h}_k|^2  \right)  \right]  \label{Laplace_I_r_(a0)} \\
	& = \mathbb{E}_{\hat{\Phi}}\left[ \prod_{k \in \hat{\Phi}} \mathbb{E}_{ \hat{h}}\left[ \exp\left(- s d_{k, \, 0}^{-\alpha} |\hat{h}_k|^2  \right)  \right]   \right] \label{Laplace_I_r_(a)}       \\
	& = \mathbb{E}_{\hat{\Phi}}\left[ \prod_{j \in \hat{\Phi}} \frac{1}{1 +  s\mu d_{j, \, 0}^{-\alpha}} \right] \label{Laplace_I_r_I}\\
	& = \exp \left( -2\lambda' \pi \int_{d}^{\infty} \frac{ s \mu x^{-\alpha+1}}{1 +  s\mu x^{-\alpha}} \, {\rm d}x   \right) \label{Laplace_I_r_(b)} \\
	& = \exp \left( \frac{2 \mu \lambda' \pi s d^{2-\alpha} }{2-\alpha} \pFq{2}{1}{1, 1-\frac{2}{\alpha}}{2-\frac{2}{\alpha}}{- \mu d^{-\alpha} s} \right), \label{Laplace_I_r}
\end{align}
where $\hat{\Phi}$ is a thinning process of $\Phi \setminus \Phi_0$ with intensity $\lambda' = \lambda/3$, and where \eqref{Laplace_I_r_(a)} follows from the fact that $\hat{h}_k$ are i.i.d. for all $k \in \hat{\Phi}$, \eqref{Laplace_I_r_I} is due to the fact that $| \hat{h}_k |^2 \sim \exp(\mu)$, and \eqref{Laplace_I_r_(b)} is based on the probability generating functional of the underlying PPP \cite{Haenggi12}.

With the obtained PDFs of $d$ and $U$ shown in \eqref{Eq_PDF_d} and \eqref{Eq_PDF_U}, respectively, and the Laplace transform of $I_1$ given by \eqref{Laplace_I_r}, the coverage probability of a typical UE can be analyzed.
\subsection{Coverage Probability}
\label{Section-CoverageProbability}
Now, we are in a position to formalize the coverage probability of a typical UE in the following theorem.
\begin{theorem}[$\ell_1$-Toeplitz matrix form of the coverage probability] \label{Theorem_TypeI}
	Given that joint transmission is applied to BSs in the cooperation set of a typical UE, with a prescribed outage threshold $\gamma$, the coverage probability of a typical UE can be calculated as
	\begin{equation} \label{Eq-CP-CaseI}
		\mathcal{P}_{1}(\gamma, \lambda, \alpha)=\int_{x>0} f_d(x) \left \| \exp(\bm{Q}(d))  \right \|_1 {\mathrm{d}}x,
	\end{equation}
	where $f_d(x)$ is shown in \eqref{Eq_PDF_d}, and $\bm{Q}(d)$ is an $m \times m$ lower triangular Toeplitz matrix, expressed as
	\begin{equation}
		\bm{Q}(d) = \left[ \begin{matrix}
				q_0 \\ q_1 & q_0 \\ q_2 & q_1 & q_0 \\ \vdots & \vdots & & \ddots \\ q_{m - 1} & \cdots & q_2 & q_1 & q_0
					\end{matrix} \right],
	\end{equation}
	with the entry $q_n$ given by
	\begin{align} \label{Eq_q_n}
		q_n &= \lambda' \pi d^2 \delta(n)	-\lambda' \pi d^2 \,\frac{2}{2-n\alpha} \left(\frac{1}{\Omega} m \mu \right)^n \gamma^n \nonumber\\
		& \quad \times{} \pFq{2}{1}{n+1, n-\frac{2}{\alpha}}{n+1-\frac{2}{\alpha}}{-\frac{1}{\Omega} m \mu \gamma}.
	\end{align}
\end{theorem}

\begin{IEEEproof}
			See Appendix~\ref{Proof_Theorem_Coverage_Vertex_b}.
\end{IEEEproof}

As an application of Theorem~\ref{Theorem_TypeI}, we consider the special case of single transmit antenna at each BS, i.e., $M=1$. In such a case, we get a simple expression as summarized in the following corollary.

\begin{corollary}\label{Corollary_TypeI}
In the case of $M =1$, the coverage probability given by \eqref{Eq-CP-CaseI} reduces to
\begin{align}
	\mathcal{P}_{1}(\gamma, \alpha)
		&=  \frac{1}{\left(1+\frac{V_1(\gamma)}{3}\right)^2} - \frac{2m\gamma V_2(\gamma)}{3\Omega\left(1+\frac{V_1(\gamma)}{3}\right)^3} \nonumber\\
		 &\quad  {}+ \frac{(m\gamma)^2 V_2(\gamma)^2}{3\Omega^2\left(1+\frac{V_1(\gamma)}{3}\right)^4}  + \frac{(m\gamma)^2V_3(\gamma)}{3\Omega^2\left(1+\frac{V_1(\gamma)}{3}\right)^3}, \label{Eq-CoMP-Rayleigh}
\end{align}
where
\begin{align}
	V_1(\gamma) & \triangleq  \frac{2 \mu m \gamma }{(\alpha-2)\Omega} \, \pFq{2}{1}{1, 1-\frac{2}{\alpha}}{2-\frac{2}{\alpha}}{-\frac{\mu m\gamma }{\Omega}}, \label{Eq-V1} \\
	V_2(\gamma) & \triangleq \frac{2 \mu }{2-\alpha} \, \pFq{2}{1}{1, 1-\frac{2}{\alpha}}{2-\frac{2}{\alpha}}{-\frac{\mu m\gamma }{\Omega}} \nonumber\\
		& \quad {}+ \frac{\mu^2 m \gamma}{(\alpha-1)\Omega} \, \pFq{2}{1}{2, 2-\frac{2}{\alpha}}{3-\frac{2}{\alpha}}{-\frac{\mu m\gamma }{\Omega}}, \label{Eq-V2} \\
	V_3(\gamma) & \triangleq \frac{2\mu^2}{\alpha -1} \, \pFq{2}{1}{2, 2-\frac{2}{\alpha}}{3-\frac{2}{\alpha}}{-\frac{\mu m\gamma }{\Omega}} \nonumber\\
		& \quad {}- \frac{4 \mu^3 m\gamma }{(3\alpha -2)\Omega} \, \pFq{2}{1}{3, 3-\frac{2}{\alpha}}{4-\frac{2}{\alpha}}{-\frac{\mu m\gamma }{\Omega}}. \label{Eq-V3}
\end{align}
\end{corollary}
\begin{IEEEproof}
     See Appendix \ref{Appendix-A}.
\end{IEEEproof}

Notice that \eqref{Eq-CoMP-Rayleigh} demonstrates that the coverage probability in the case of $M=1$ is independent of the intensity of the BSs ($\lambda$). By using a similar approach as described above, it is not hard to show that this conclusion holds as well even if $M > 1$. To sum up, this means that increasing the number of BSs will not benefit the coverage probability. An intuitive interpretation of this conclusion is that an increase in desired signal power is exactly counter-balanced by that in unwanted interference power. This conclusion agrees with empirical observations in interference-limited urban networks \cite{AndrewsTCOM1111}.
\subsection{Spectral Efficiency} \label{Spectral_efficiency_JT}
By using a similar method to \cite[Eq. (12)]{8314102}, the spectral efficiency of the JT scheme can be approximated as
\begin{align}
	\tau_1(\alpha) 
	& \approx \int\limits_{x>0} \mathbb{E}\left[\ln \left(1+  \frac{ d^{-\alpha}{\mathbb{E}[U_1]} }{\sum\limits_{k \in \hat{\Phi}} d_{k, \, 0}^{-\alpha}\mathbb{E}\left[ \left|  \hat{h}_k\right|^2 \right]}\right) \bigg| d \right]  f_d(x) {\mathrm d} x \label{Eq-Spectral_1-a} \\
	&= \int\limits_{x>0} \mathbb{E}\left[\ln \left(1+  \frac{\frac{\Omega}{\mu}} {\sum\limits_{k \in \hat{\Phi}} d_{k, \, 0}^{-\alpha}  d^{\alpha}  }\right) \bigg| d \right] f_d(x) {\mathrm d} x \nonumber\\
	&= \int\limits_{s>0} \int\limits_{x>0} \frac{1}{s} \left[1- \exp\left(-\frac{s \Omega}{\mu} \right) \right]
	\exp\left\{-\lambda'\pi d^2 \right. \nonumber\\
	& \quad \left. \times {} \left[\exp(-s) + s^{\frac{2}{\alpha}}\gamma\left(1-\frac{2}{\alpha}, s\right) \right]-1 \right\} {\mathrm{d}s} \, f_d(x) \,{\mathrm{d}x} \label{Eq-Spectral_1-b}\\
	&= 9 \int\limits_{s>0} \frac{1}{s} \left(1- \exp\left(-\frac{s \Omega}{\mu} \right)\right)  \nonumber\\
	&\quad \times \left[\exp(-s)+ s^{\frac{2}{\alpha}}\gamma\left(1-\frac{2}{\alpha}, s\right)+2 \right]^{-2}  {\mathrm{d}s}, \label{Eq-Spectral_1-c}
\end{align}
where \eqref{Eq-Spectral_1-b} is derived by using the lemma in \cite{Hamdi2010A}, and \eqref{Eq-Spectral_1-c} is obtained by substituting \eqref{Eq_PDF_d} into \eqref{Eq-Spectral_1-b} as well as performing some basic calculus.

For comparison purposes, the exact expression for the spectral efficiency is derived as
 \begin{align}
	\tau_1(\alpha)
		& \triangleq \mathbb{E}\left(\ln \left(1 + \Gamma_1\right)   \right) \nonumber\\
		&= \int_{x>0}\int_{t>0} \mathcal{P}\left[\ln \left(1+  \frac{d^{-\alpha} U}{I_1}\right) > t \right] {\rm d}t \, f_d(x) {\rm d}x \nonumber\\
		&=  \int_{x > 0}\int_{t>0}  \sum_{k=0}^{m-1}\frac{1}{k!}\left(-\frac{m}{\Omega}d^{\alpha}\left(\exp(t)-1\right)\right)^k \nonumber\\
		&\quad \times{} \frac{\partial^k L_{I_1}(s)}{\partial s^k} \bigg|_{s=\frac{m}				{\Omega}d^{\alpha}\left(e^t-1\right)} \,{\rm d}t \, f_d(x) {\rm d}x\label{Eq-Spectral_1_a}\\
		&= \int_{x > 0}\int_{t>0} \left \| \exp(\bm{Q}(d)) \right\|_1\bigg|_{\gamma = e^t - 1} \,{\rm d}t \, f_d(x) {\rm d}x. \label{Eq-Spectral_1_b}
 \end{align}
In the special case of $M=1$, we have $\Omega=3$ according to \eqref{Eq_PDF_T1-Approx-2a}. Then, assuming the path-loss exponent $\alpha = 4$, the spectral efficiency of the proposed JT scheme can be numerically calculated as per \eqref{Eq-Spectral_1_b}, yielding
\begin{equation}  \label{Spectral_JT}
	\tau_1(\alpha = 4) =  2.24 \text{ nats/sec/Hz}.
\end{equation}

\begin{remark}[Performance analysis of Types II and III UEs] Notice that the performance analysis developed in this section for Type I UEs exploits the fact that the typical UE is equidistant from three serving BSs, as shown in Eq.~\eqref{Eq_RxSINR-CaseI}. As far as Types II and III UEs are concerned, the typical UE has different distances to three serving BSs and, thus, the received SIR given by Eq.~\eqref{Eq_RxSINR} cannot be reduced to Eq.~\eqref{Eq_RxSINR-CaseI}, and the subsequent analyses of coverage probability and spectral efficiency cannot be repeated in a similar way. As an alternative, some advanced technique needs to be developed to attain the distribution function of the desired signal expressed by the numerator of Eq.~\eqref{Eq_RxSINR}. This is beyond the scope of this paper and will be tackled in our future work.
\end{remark}

\section{Dynamic Point Selection/Muting}	
\label{Section-DPSM}
Although the JT scheme described above benefits lower outage probability, it requires all BSs in the cooperation set to simultaneously serve a target UE, leading to higher hardware and coordination costs. To get a tradeoff between the higher costs and lower outage probability, the technique of dynamic point selection/muting can be applied \cite{8115171, 7008426}. Specifically, not all BSs in the cooperation set but only the one with the best channel quality (i.e., the product of the large-scale path loss and small-scale fading) is chosen to serve the target UE while the remaining BSs keep silent. This scheme is called {\it optimal point selection} (OPS) in the sequel.

\subsection{Optimal Point Selection}
\label{Section-OptimalPointSelection}
This subsection derives the coverage probability and spectral efficiency of the OPS scheme in sequence. To start with, the aggregate interference at a typical UE, denoted $I_2$, comes from all BSs in the complement set $\Phi \setminus \Phi_0$, and is given by
\begin{equation}
	I_2 = \sum_{k \in \Phi \setminus \Phi_0}d_{k, 0}^{-\alpha} ~ \left|\bm{h}_{k, \, 0}^H \, \frac{\bm{h}_{k}}{\|{\bm{h}_k\|}} \right|^2 =\sum_{k \in \Phi \setminus \Phi_0}d_{k, 0}^{-\alpha} ~ g_k, \label{L_I_2_a}
\end{equation}
where $g_k \triangleq \left| \bm{h}_{k, \, 0}^H \, \frac{\bm{h}_{k}}{\|{\bm{h}_k\|}} \right|^2$ is of exponential distribution with unit mean. Then, the Laplace transform of $I_2$ can be derived as
\begin{align}
	L_{I_2}(s)
		& = \mathbb{E}_{\Phi, g_j}\left[ \exp\left(- s \sum_{j \in \Phi \setminus \Phi_0} g_j d_{j, \, 0}^{-\alpha} \right)  \right] \label{Laplace_I_2_a} \\
		& = \exp \left( \frac{2\lambda \pi s d^{2-\alpha} }{2-\alpha} \pFq{2}{1}{1, 1-\frac{2}{\alpha}}{2-\frac{2}{\alpha}}{-d^{-\alpha} s} \right). \label{Laplace_I_2}
\end{align}
\subsubsection{Coverage Probability}
By jointly applying the theories of order statistics and stochastic geometry, the coverage probability of a typical UE in case the OPS is applied, can be formalized as follows.
\begin{theorem}\label{Theorem_TypeI_2}
	Given that the optimal point selection technique is applied to BSs in the cooperation set of a typical UE, with a prescribed outage threshold $\gamma$, the coverage probability of a typical UE can be calculated as
	\begin{align}  \label{Coverage_CoMp_OneBase}
		& \mathcal{P}_{2} (\gamma, \lambda, \alpha)
			  = \int_{x > 0} f_d(x) \left\{ 3\left \| \exp(\bm{Q}'(d)) \right\|_1 \right. \nonumber\\
			&\quad \left. -{} \mathbb{E}_{I_2} \left[ 3Q^2\left(M, \gamma d^{\alpha}I_2 |d\right) - Q^3\left(M, \gamma d^{\alpha} I_2 |d\right) \right] \right\} {\rm d}x,
	\end{align}
	where $\bm{Q}'(d)$ is an $M \times M$ lower triangular Toeplitz matrix, expressed as
	\begin{equation} \label{Matrix_Q'}
		\bm{Q}'(d) = \left[ \begin{matrix}
					q_0' \\ q_1' & q_0' \\ q_2' & q_1' & q_0' \\ \vdots & \vdots & & \ddots \\ q_{M-1}' & \cdots & q_2' & q_1' & q_0'
				\end{matrix}\right],
	\end{equation}
	with the entry $q_n'$ given by
	\begin{equation} \label{Eq_q_n_2}
		q_n' = \lambda \pi d^2 \delta(n)	- \frac{2\lambda \pi d^2}{2-n\alpha}\,\gamma^n \pFq{2}{1}{n+1, n-\frac{2}{\alpha}}{n+1-\frac{2}{\alpha}}{-\gamma},
	\end{equation}
	 and where
	\begin{align}
    	         \lefteqn{\mathbb{E}_{I_2} \left[Q^n  \left(M, \gamma d^{\alpha} I_2 |d \right) \right]} \nonumber \\
		 	& = \left(\sum_{k=0}^{M - 1}\frac{1}{k!}\left(-\gamma d^{\alpha} \right)^k \frac{\partial^k L_{I_2}(s)}{\partial s^k}\right)^n \bigg|_{s = \gamma d^{\alpha}},\ n = 2, 3.
	\end{align}
\end{theorem}

\begin{IEEEproof}
	See Appendix \ref{Appendix-B}.
\end{IEEEproof}

As an application of Theorem~\ref{Theorem_TypeI_2}, we consider the special case with $M=1$, i.e., there is only a single transmit antenna at each BS. Recalling that $Q(1, x) = \exp(-x)$,  \eqref{Max_G} in Appendix~\ref{Appendix-B} reduces to
 \begin{equation}\label{Eq-MAx-Co}
 {\rm{Pr}} \left[ G > \gamma d^{\alpha} I_2 |d \right]
 = 3L_{I_2}\left( \gamma d^{\alpha}  \right) - 3L_{I_2}\left( 2\gamma d^{\alpha} \right) + L_{I_2}\left( 3 \gamma d^{\alpha} \right).
 \end{equation}
On the other hand, by virtue of \eqref{Laplace_I_2}, performing some basic calculus yields the coverage probability:
\begin{align} \label{Eq-CoMP-OneBase_Rayleigh}
	\mathcal{P}_{2} (\gamma, \lambda, \alpha)
		&= 3\left(1+V_4(\gamma)\right)^{-2} - 3\left(1+ V_4(2\gamma)\right)^{-2} \nonumber\\
		&\quad {}+  \left(1+ V_4(3\gamma)\right)^{-2},
\end{align}		
where
\begin{equation}\label{Eq-K}
	 V_4(\gamma) \triangleq \frac{2 \gamma}{\alpha-2} \, \pFq{2}{1}{1, 1-\frac{2}{\alpha}}{2-\frac{2}{\alpha}}{-\gamma}.
\end{equation}

Like \eqref{Eq-CoMP-Rayleigh}, \eqref{Eq-CoMP-OneBase_Rayleigh} also demonstrates that the coverage probability in the case of dynamic point selection/muting with $M = 1$ is independent of the intensity of BSs (i.e., $\lambda$). By using a similar approach as above, this conclusion can be shown to hold as well even if $M > 1$.
\subsubsection{Spectral Efficiency}
By using a similar approach to \eqref{Eq-Spectral_1-a}-\eqref{Eq-Spectral_1-c}, the spectral efficiency of the OPS scheme can be approximated as
\begin{equation}  \label{Spectral_2_Approx}
	\tau_{2} (\alpha) \approx  \int_{s>0} \frac{1}{s} \frac{1- \exp\left(-sN(M)\right)}{ \left[\exp(-s) + s^{\frac{2}{\alpha}}\gamma\left(1-\frac{2}{\alpha}, s\right) \right]^2}  {\mathrm{d}s},
 \end{equation}
where $N(M) \triangleq \int_{0}^{1}2u^2 x(u) {\mathrm{d}}u$, with $x(u)={F^{-1}}_{g_i'}(u)$ and $F_{g_i'}(x)$ given by \eqref{Eq-CDF_G}.

 By definition, the exact spectral efficiency of the OPS scheme can be derived and given by
 \begin{align}  \label{Spectral_Efficiency_OPS}
 \tau_2 (\alpha)
	&= \int\limits_{t>0} \int\limits_{x>0} f_d(x) \left\{ 3 \left \| \exp(\bm{Q}'(d)) \right\|_1 \right. \nonumber\\
	& \left. {}- \mathbb{E}_{I_2}\left[ 3Q^2\left(M, \gamma d^{\alpha}I_2 |d\right) - Q^3\left(M, \gamma d^{\alpha} I_2 |d \right) \right] \right\} {\rm d}x \, {\rm d}t.
 \end{align}
 where $\gamma = e^t -1$. In the case of $M = 1$ and $\alpha = 4$, the spectral efficiency of the OPS scheme, numerically computed as per \eqref{Spectral_Efficiency_OPS}, is
\begin{equation} \label{Spectral_OPS}
	\tau_2(\alpha = 4) = 1.03 \text{ nats/sec/Hz}.
\end{equation}

For comparison purposes, next a random point selection (RPS) scheme is discussed, where one BS in the cooperation set is randomly chosen to serve the target UE while the other BSs serve other UEs at the same time. Compared with the OPS scheme descried above, RPS has higher resource utilization, yet with lower spectral efficiency, as shown below.

\subsection{Random Point Selection}
In this case, a typical UE is randomly served by only one BS without CoMP. Without loss of generality, the serving BS is denoted $A_0$. Unlike the preceding JT and OPS cases, the aggregate interference at a typical UE, denoted $I_3$, comes from all BSs in the complement set of $\Phi$, i.e., $\Phi \setminus \{A_0\} $. Mathematically speaking, we have
\begin{equation}
    	I_3 = \sum_{ k \in \Phi \setminus \{A_0\} } d_{k, \, 0}^{-\alpha}\, g_k,
\end{equation}
where $g_k$ is as defined right after \eqref{L_I_2_a}. The Laplace transform of $I_3$ can be readily computed as
\begin{align}
    L_{I_3}(s)
		 &= \frac{1} {\left(1 + \frac{s} {d^{\alpha}}\right)^{2}} \mathbb{E}_{\Phi}\left[\prod_{ k \in \Phi \setminus \{A_0\}} \mathbb{E}_{g_k} \left[\exp\left(- s  g_k  d_{k, 0}^{-\alpha} \right)\right]\right]  \nonumber  \\
		&= \left(1 + \frac{s} {d^{\alpha}}\right)^{-2} \hspace{-8pt}\exp\left(\frac{2\lambda \pi s d^{2-\alpha} }{2-\alpha} \pFq{2}{1}{1, 1-\frac{2}{\alpha}}{2-\frac{2}{\alpha}}{-\frac{s}{d^{\alpha}}}\right). \label{Laplace_I'_r}
\end{align}

\subsubsection{Coverage Probability}
 By using the Alzer's lemma in \cite{Alzer1997On}, the coverage probability of a typical UE in the case of RPS can be derived, as formalized below.
\begin{theorem} \label{Theorem_TypeI_3}
	The coverage probability of a typical UE in the case of random point selection is upper bounded by
	\begin{align}
		\mathcal{P}_{3} (\gamma, \lambda, \alpha)
			&= \int_{x>0} {f_d(x) \sum_{k=0}^{M-1} \frac{(-\gamma d^\alpha)^k}{k!} \frac{\partial^k L_{I_3}(s)}{\partial s^k} \bigg|_{s=\gamma d^\alpha}}  {\rm d}x \label{Coverage_NoCoMP_a} \\
			&\leq \sum\limits_{k=1}^{M}(-1)^{k+1} {M \choose k} \int_{x>0}   f_{d}(x) L_{I_3}\left(k \beta \gamma d^{\alpha} \right)   {\rm d}x, \label{Coverage_NoCoMP_b}
	\end{align}
	where $\beta = (M!)^{-1/M}$.
\end{theorem}
			
In particular, if single transmit antenna is deployed at each BS ($M=1$) and Rayleigh fading is assumed, \eqref{Coverage_NoCoMP_a} reduces to
\begin{align}
	\mathcal{P}_{3} (\gamma, \lambda, \alpha)
		& =  \int_{x > 0} f_d(x) \, L_{I_3}(\gamma d^{\alpha}) \, {\rm d}x \nonumber\\
	       &=  \left[(1+\gamma) (1 + V_4(\gamma))\right]^{-2},  \label{Coverage_NoCoMP-b}
\end{align}
where $V_4$ is previously defined in \eqref{Eq-K}. Like \eqref{Eq-CoMP-Rayleigh}, \eqref{Coverage_NoCoMP-b} also demonstrates that the coverage probability of a typical UE is independent of the intensity of BSs, i.e., $\lambda$.
\subsubsection{Spectral Efficiency}
Using a similar approach as in Section~\ref{Spectral_efficiency_JT}, the spectral efficiency of the RPS scheme can be approximated as follows	
\begin{equation}
	\tau_{3} (\alpha) \approx  \int_{s>0} \frac{1}{s} \frac{\left(1- \exp\left(-sM\right) \right) \exp(-2s)} { \left[\exp(-s) + s^{\frac{2}{\alpha}}\gamma\left(1-\frac{2}{\alpha}, s\right) \right]^2} \, {\mathrm{d}s}. \label{Spectral_3_Approx}
\end{equation}
On the other hand, by definition, the exact spectral efficiency of the RPS scheme can be computed as
\begin{align} \label{Spectral_RPS}
	\tau_{3} (\alpha) &= \int_{t > 0}\int_{x > 0} f_d(x) \sum_{k=0}^{M-1} \frac{(-(\exp(t)-1) d^\alpha)^k}{k!} \nonumber\\
	&\quad \times{} \frac{\partial^k L_{I_3}(s)}{\partial s^k} \bigg|_{s=(e^t -1) d^\alpha}  {\rm d}x \, {\rm d}t.
\end{align}
In the case of $M = 1$ and $\alpha = 4$, the spectral efficiency of the RPS scheme, numerically computed according to \eqref{Spectral_RPS}, is
\begin{equation} 
\label{Spectral_RPS_M1}
	\tau_3(\alpha = 4) = 0.27  \text{ nats/sec/Hz}.
\end{equation}

To sum up, with the obtained \eqref{Spectral_JT}, \eqref{Spectral_OPS}, and \eqref{Spectral_RPS_M1}, corresponding to the spectral efficiencies of the JT, OPS, and RPS schemes, respectively, it is obvious that the JT scheme attains the highest spectral efficiency whereas the RPS scheme yields the lowest. This result is not surprising since increasing coordination costs benefit higher spectral efficiency.
\section{Simulation Results and Discussions}
\label{Section-Simulation}
In this section, numerical results computed as per the previously obtained analytical expressions are presented and discussed, in comparison with extensive Monte-Carlo simulation results. In the simulation experiments, a cellular network with a coverage area of $10^4 \times 10^4$ squared meters is considered, where the path-loss exponent and BS intensity are set to $\alpha =4$ and $\lambda = 0.02$, respectively. The channel fading from each transmit antenna at BSs to a typical UE is subject to Rayleigh fading. Moreover, for the case with single transmit antenna at each BS, i.e., $M = 1$, according to \eqref{Eq_PDF_T1-Approx-2a}-\eqref{Eq-Appro_Gamma_Moment} and after some algebraic calculations, we have $\Omega = 7.7$ and $m = 3$. Similarly, $\Omega=16.6$ and $m = 6$ in the case of $M = 2$.

\subsection{Coverage Probability}
\begin{figure}[!tt]
	\centering
	\includegraphics[width=3.25in, clip]{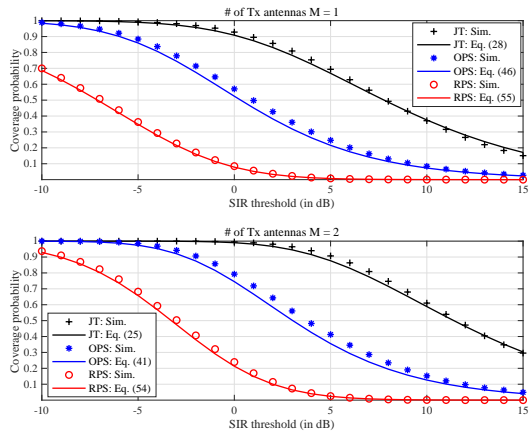}
	\caption{Coverage probabilities versus SIR threshold, with three different transmission schemes (JT: joint transmission, OPS: optimal point selection, and RPS: random point selection).}
	\label{Fig-6}
\end{figure}

Figure~\ref{Fig-6} shows the coverage probability of a typical UE versus the outage threshold $\gamma$, where the top panel corresponds to the case of $M=1$ while the bottom panel to the case of $M = 2$. For comparison purposes, the coverage probabilities of three different transmission schemes, namely, JT, OPS, and RPS, are plotted. For a particular outage threshold value, it is seen that the JT has the highest coverage probability whereas the RPS gets the lowest, as expected. On the other hand, taking the JT for instance, it is observed that the numerical results computed as per \eqref{Eq-CoMP-Rayleigh} are slightly smaller than the corresponding simulation results. Similar observations can be made in the OPS and RPS cases, as shown in Fig.~\ref{Fig-6}. In other words, the coverage probability of a typical UE is slightly underestimated in the preceding analyses. The reason behind this interesting observation is far more complex than what seems at first sight. Specifically, although \eqref{Eq_PDF_T1-Approx} is an approximate PDF, it is accurate and has little effect on the derived coverage probability as previously discussed in Remark~\ref{Remark_2} at the end of Section~\ref{Section-JT-A-b}. The approximation given by \eqref{Eq_L_1_b} is also accurate, as previously illustrated in Fig.~\ref{Fig-5}.

In fact, this underestimation is introduced by the assumption of independence between BSs and a typical UE. More concrete evidence is provided below. By recalling the Slivnyak-Mecke theorem in stochastic geometry \cite[p. 132]{Chiu13}, a typical UE can be assumed to be located at the origin $(0, 0) \in \mathbb{R}^2$, without loss of generality. This implies that the location of a typical UE is independent of the locations of BSs. However, as far as the worst-case UEs under study in this paper is concerned, a typical UE has the same distance from its three nearest BSs. Clearly, the location of a typical UE in our work is dependent upon the locations of BSs. For better clarity, let us take a close look at the aggregate signal power at the origin, defined as
\begin{equation}  \label{Eq-S1}
S_1 \triangleq \sum_{k \in \Phi}  d_{k, \, 0}^{-\alpha} \, g_k,
\end{equation}
where $g_k$ is defined immediately after \eqref{L_I_2_a}. It is well-known that $S_1$ is subject to a skewed stable distribution \cite{WinPROCIEEE0902}, whose probability densities exist but, with a few exceptions, they are not known in closed form. In particular, if $\alpha = 4$, the PDF of $S_1$ is of the L\'{e}vy type, explicitly expressed as \cite{GeTWC1110}
\begin{equation} \label{Eq-InterferenceT}
	f_{S_1}(x) = \frac{\lambda}{4}\left(\frac{\pi}{x}\right)^{\frac{3}{2}}\exp\left(-\frac{\lambda^2 \pi^4}{16 x}\right).
\end{equation}

\begin{figure}[t]
		\centering
		\includegraphics[width=3.25in, clip]{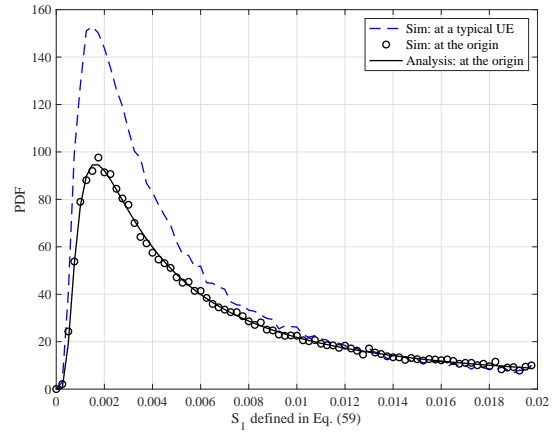}
		\caption{PDFs of the aggregate signal defined in Eq.~\eqref{Eq-S1} at a typical UE and at the origin.}
		\label{Fig-7}
\end{figure}

Figure~\ref{Fig-7} shows the simulation results of the aggregate signal power at the origin and at a typical UE which has the same distance to its nearest three BSs, compared with the  numerical results computed with \eqref{Eq-InterferenceT}, given the BS intensity $\lambda = 0.02$. It is seen that the former simulation results accord fully with the numerical results whereas the latter simulation results deviate from the numerical ones significantly. In particular, the PDF of the aggregate signal power at a typical UE has higher kurtosis than that of the power at the origin. This means that the former has infrequent extreme deviations or, equivalently, this reflects the dependence of different signals transmitted from BSs to a typical UE.

\begin{figure}[!t]
	\centering
	\includegraphics[width=3.0in, clip]{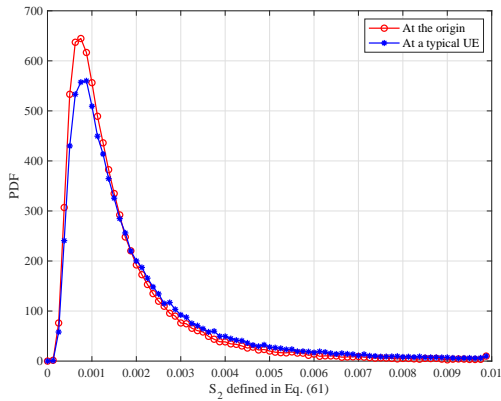}
	\caption{PDFs of the interference defined in Eq.~\eqref{Eq-S2} at the origin and at a typical UE.}
	\label{Fig-8}
\end{figure}

As far as the JT scheme under study is concerned, the interference can be expressed as
\begin{equation}  \label{Eq-S2}
	S_2 \triangleq \sum_{k \in \Phi \setminus \Phi_0}  d_{k, \, 0}^{-\alpha}\, g_k.
\end{equation}
Figure~\ref{Fig-8} shows the simulated PDFs of the powers of $S_2$ at the origin and at a typical UE which has the same distance from its three nearest BSs, respectively (exact PDF of $S_2$ is not mathematically tractable, to the best of the authors' knowledge). Clearly, the difference between these two PDF curves is much smaller than that in Fig.~\ref{Fig-7}. As a consequence, we may conclude that the difference between the powers of $S_1$ received at the origin and at a typical UE under study in this paper is mainly caused by the signals from the nearest three BSs. This is indeed the reason why the obtained analytical results has led to a slight underestimation of the coverage probability, as shown in Fig.~\ref{Fig-6}.

\subsection{Spectral Efficiency}
\begin{figure}[!t]
	\centering
	\includegraphics[width=3.0in, clip]{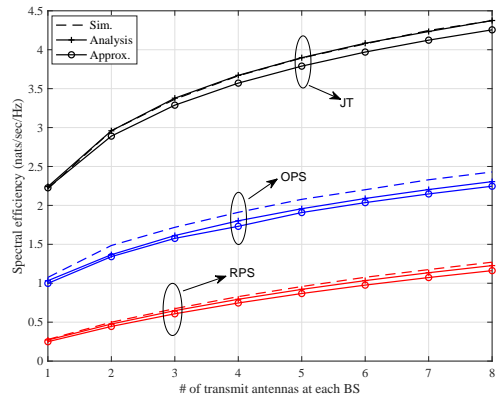}
	\caption{Spectral efficiency versus the number of transmit antennas at each BS (i.e., $M$) (JT: \eqref{Eq-Spectral_1_b} vs. \eqref{Eq-Spectral_1-c}; OPS:  \eqref{Spectral_Efficiency_OPS} vs. \eqref{Spectral_2_Approx}; RPS: \eqref{Spectral_RPS} vs. \eqref{Spectral_3_Approx}).}
	\label{Fig-9}
\end{figure}

Figure~\ref{Fig-9} compares the spectral efficiencies of the JT, OPS and RPS schemes. As expected, the spectral efficiency of all schemes increases with larger number of transmit antennas at BS (i.e., $M$). For a fixed $M$, the JT scheme has the highest spectral efficiency while the RPS gets the lowest, since the former requires higher cooperation and hardware costs. Importantly, it is observed from Fig.~\ref{Fig-9} that the numerical results pertaining to the JT and RPS schemes agree very well with the simulation results, whereas those of the OPS scheme underestimate the simulation ones. This observation implies that the dependence discussed in the previous subsection has little effect on the accurate analysis of the spectral efficiencies of the JT and RPS schemes. Also, it is seen that the approximated numerical results match well with the exact analytical results, which illustrates the effectiveness of the analysis.
\subsection{Poisson-Delaunay Triangulation vs. Poisson-Voronoi Tessellation}

\subsubsection{Comparison with Poisson-Voronoi tessellation without CoMP}
\label{Subsection_NoCoMP}

To illustrate the effectiveness of the Poisson-Delaunay triangulation based JT scheme, besides the performance analysis of Type~I UEs, this subsection shows the simulation results pertaining to Types~II and III UEs defined in Section~\ref{Subsect_TypesUEs}, in comparison with their counterparts in the conventional cellular systems based on Poisson-Voronoi tessellation.

\begin{figure}[!t]
	\centering
	\includegraphics[width=3.0in, clip]{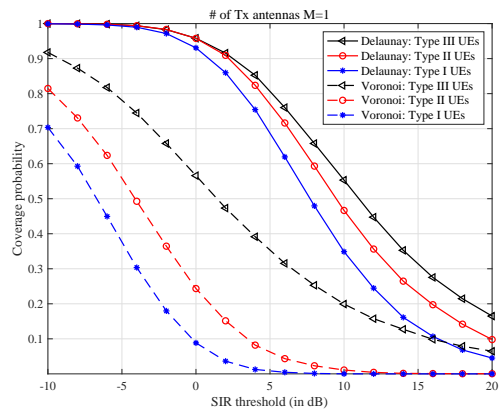}
	\caption{Coverage probabilities of different types of UEs versus the SIR threshold, with $M = 1$.}
	\label{Fig-10}
\end{figure}

Figure~\ref{Fig-10} illustrates the coverage probabilities of three types of UEs, compared with their counterparts in the dual Poisson-Voronoi tessellation. It is seen that, in the conventional Poisson-Voronoi scenario, the performance of Type~I UEs is the worst and Type~II UEs perform worse than Type~III UEs. The reason is that they have shorter and shorter distances from the serving BS by noting that Type~I UEs are at the vertices of each Poisson-Voronoi cell, Type~II UEs are on the edge of each cell, and Type~III UEs are inside each cell (cf. Fig.~\ref{Fig-2}). In the scenario of Poisson-Delaunay triangulation, similar observations can be made, namely, Type~I UEs perform the worst since the strength of desired signals received at Type~I UEs is the lowest whereas the strengths of interfering signals on them are almost identical. More specifically, if we consider only the path-loss effect, by recalling the inequality of arithmetic and geometric means, we have $d_1^{-\alpha} + d_2^{-\alpha} + d_3^{-\alpha} \ge 3 \sqrt[3]{(d_1 d_2 d_3)^{-\alpha}}$, where the equality holds if and only if $d_1 = d_2 = d_3$ (corresponding to Type~I UEs). Finally, Fig.~\ref{Fig-10} shows that all UEs in the scenario of Poisson-Delaunay triangulation significantly outperform their counterparts in the conventional Poisson-Voronoi scenario, indicating the effectiveness of the proposed triangulation scheme. Clearly, this performance gain comes from the cooperation of BSs. Next, we will demonstrate the interference mitigation capability of our triangulation scheme.
\subsubsection{Comparison with Poisson-Voronoi tessellation with dynamic cooperation set}
For fairness, we compare the triangulation scheme with the conventional Poisson-Voronoi scheme with three dynamic cooperating BSs, in terms of spectral efficiency. Since the three serving BSs of a Type~I UE in the triangulation scheme are exactly the nearest ones, the desired signal powers under these two schemes are identical, say $d$. Then, by recalling \eqref{Eq_PDF_U}, the MGF of the desired signal conditioned on $d$ can be computed as
\begin{equation}
	M_{S} = \left(1+ \frac{\Omega}{m} d^{-\alpha} z \right)^{-m}. \label{Eq_Ms}
\end{equation}

Now, we compare their interference powers $I_1$ and $I_2$ given by \eqref{Eq_I} and \eqref{L_I_2_a}, respectively. By inserting $\mu = 3$ into \eqref{Laplace_I_r_I}, for a given $d$, the MGF of $I_1$ can be readily shown as
\begin{equation}\label{Eq_MI_a}
	M_{I_1} = \mathbb{E}_{\hat{\Phi}} \left[\prod_{k \in \hat{\Phi}} \left(1+3 z d_{k,0}^{-\alpha} \right)^{-1}  \right].
\end{equation}
 Likewise, the MGF of $I_2$ given by \eqref{L_I_2_a} can be derived as
\begin{align}
	M_{I_2} &= \mathbb{E}\left[\exp\left(-z \sum_{k \in \Phi \setminus \Phi_0} d_{k,0}^{-\alpha} g_k \right) \right] \nonumber  \\
		     &\approx \mathbb{E}\left[\exp\left(-z \sum\limits_{j =1}^\infty d_{j, \, 0}^{-\alpha} \sum\limits_{k \in \Phi_j} g_k  \right) \right] \label{Eq_M2_a1}  \\
		     &= \mathbb{E}_{\hat{\Phi}} \left[\prod_{k \in \hat{\Phi}} \mathbb{E}\left[\exp\left(-z d_{k,0}^{-\alpha} \hat{g}_k  \right)  \right]   \right] \label{Eq_M2_a2}  \\
		     &= \mathbb{E}_{\hat{\Phi}} \left[\prod_{k \in \hat{\Phi}} \left(1+z d_{k,0}^{-\alpha} \right)^{-3}  \right], \label{Eq_M2_a}
\end{align}
where $\hat{\Phi}$ is a thinning process of $\Phi \setminus \Phi_0$ with intensity $\lambda' = \lambda/3$, and $\hat{g}_k \triangleq \sum_{k \in \Phi_j} g_k$, which follows a Gamma distribution with shape parameter $3$ and unit scale factor.

\begin{figure}[!t]
	\centering
	\includegraphics [width=3.0in, clip, keepaspectratio]{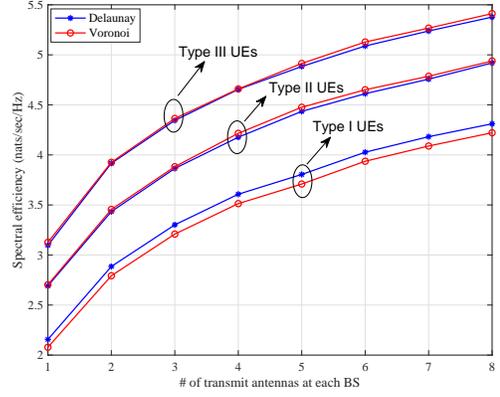}
	\caption{Spectral efficiency for different types of UEs simulated under Poisson-Delaunay triangulation based JT scheme and Poisson-Voronoi tessellation with three dynamic cooperating BSs.}
	\label{Fig_11}
\end{figure}

To derive the spectral efficiency, we exploit the lemma reported in \cite{Hamdi2010A}, which reads
\begin{equation}
	\ln\left(1+\frac{X}{Y}\right) = \int_{z>0} \frac{1}{z} \left(1-\exp\left(-zX\right)\right)\exp(-zY) \mathrm{d}z.
\end{equation}
Then, the spectral efficiency can be readily computed as
\begin{align}
	R &= \int_{x>0} \mathbb{E}\left[\ln\left(1+\frac{S}{I}\right)\bigg|d \right] f_d(x) \, \mathrm{d}x \label{Eq_Spectral_a} \\
		   &= \int_{x>0} f_d(x) \int_{z>0} \frac{1}{z} \left(1- M_{S}\right) {M_{I}} \, \mathrm{d}z \mathrm{d}x, \label{Eq_Spectral_b}
\end{align}
where $M_{S}$ and $M_{I}$ denote the MGFs of the desired signal power $S$ and interference power $I$, respectively. Comparing \eqref{Eq_MI_a} with \eqref{Eq_M2_a}, since $ \left(1+3 z d_{k,0}^{-\alpha} \right)^{-1} > \left(1+z d_{k,0}^{-\alpha} \right)^{-3}$ for all $z > 0$ and $d_{k,0} > 0$, we know that $M_{I_1} > M_{I_2}$. Therefore, the spectral efficiency of the proposed scheme, computed by substituting \eqref{Eq_Ms} and \eqref{Eq_MI_a} into \eqref{Eq_Spectral_b}, is larger than that of the Poisson-Voronoi scheme, computed by substituting \eqref{Eq_Ms} and \eqref{Eq_M2_a} into \eqref{Eq_Spectral_b}. Simulation results shown in Fig.~\ref{Fig_11} corroborates this analysis. For completeness of presentation, Fig.~\ref{Fig_11} also illustrates the spectral efficiency of the other two types of UEs. It is observed that the spectral efficiency of the Poisson-Voronoi scheme is slightly higher than that of the Poisson-Delaunay scheme, for either Type II UEs or Type III UEs. This is because the former can always choose the three nearest BSs through exhaustive searching.

\section{Concluding Remarks}
\label{Section-Conclusion}
This paper analyzed the performance of a novel coordinated multi-point (CoMP) transmission scheme based on Poisson-Delaunay triangulation. Using the theory of stochastic geometry, the coverage probabilities and the spectral efficiencies of the worst-case UEs pertaining to three different transmission schemes, namely, joint transmission, optimal point selection, and random point selection, were analytically derived and compared. Numerical results demonstrated the effectiveness of the performance analyses and the superiority of the proposed approach. Thanks to the simplicity of cooperation strategy and superiority of network performance, the proposed transmission scheme is promising in the emerging small-cell networks and/or heterogeneous networks, where CoMP transmission is indispensable for higher resource utilization.
\appendices

\section{Proof of Theorem~\ref{Theorem_TypeI}}
\label{Proof_Theorem_Coverage_Vertex_b}
Since a typical UE is associated with its three nearest BSs at an equivalent distance $d$, the coverage probability can be explicitly computed as
\begin{align}
	\mathcal{P}_{1}(\gamma, \lambda, \alpha)
		& =  \mathbb{E}\left[{\rm{Pr}}[\Gamma > \gamma \,|\,d]\right]  \nonumber \\
		& =  \int_{x > 0} {\rm{Pr}}\left[\frac{d^{-\alpha}U}{I_1} > \gamma \, |\, d\right] f_d(x) \,{\rm d}x \nonumber \\
		& =  \int_{x > 0} \mathbb{E}_{I_1}\left[{\rm{Pr}}\left[U > \gamma d^{\alpha}I_1 \,|\, d, I_1\right]\right] f_d(x)\, {\rm d}x \nonumber \\
		& = \int_{x > 0} f_d(x)\sum_{k=0}^{m-1}\frac{1}{k!}\left(\frac{1}{\Omega}m \gamma d^\alpha\right)^k \nonumber\\
		&\quad \times{} \mathbb{E}_{I_1}\left[I_1^k \exp\left(-\frac{1}{\Omega}m \gamma d^\alpha I_1\right)\right] {\rm d} x \nonumber\\
		& = \int_{x > 0}   f_d(x) \sum_{k=0}^{m-1}\frac{1}{k!} \left(-\frac{m \gamma d^\alpha}{\Omega}\right)^k\nonumber\\
		&\quad \times{} \frac{\partial^k L_{I_1}(s)}{\partial s^k} \bigg|_{s=\frac{1}{\Omega}m \gamma d^\alpha} {\rm d}x, \label{Eq-Coverage1}
\end{align}
where \eqref{Eq-Coverage1} follows the relationship between the moments and the Laplace transform of a RV.
	
Then, by virtue of \eqref{Laplace_I_r_(a)}, the $L_{I_1}(s)$ used in \eqref{Eq-Coverage1} can be expressed as
\begin{align}
	 L_{I_1}(s)
		& \approx  \mathbb{E}_{\hat{\Phi}}\left[ \prod_{k \in \hat{\Phi}} \mathbb{E}_{ \hat{h}}\left[ \exp\left(- s d_{k, \, 0}^{-\alpha} |\hat{h}_k|^2  \right) \right] \right]  \nonumber \\
		&= \exp\left\{-2\lambda' \pi \int_{d}^{\infty}\left(1- \mathbb{E}_{\hat{h}}\left[\exp\left(-s \hat{h} v^{-\alpha} \right) \right] v\mathrm{d}v \right) \right\}  \nonumber \\
		&= \exp\left[\eta(s) \right], \label{Eq_Eta_a0}
\end{align}
where
\begin{align}
	\eta(s) &\triangleq -2\lambda' \pi \int_{d}^{\infty}\left( 1- \mathbb{E}_{\hat{h}}\left[\exp\left(-s \hat{h} v^{-\alpha}\right)\right]  v\mathrm{d}v  \right) \nonumber\\
		&= \lambda' \pi d^2+\frac{2}{\alpha}\lambda' \pi s^{\frac{2}{\alpha}}\mathbb{E}_{\hat{h}}\left[\hat{h}^{\frac{2}{\alpha}}\gamma\left(-\frac{2}{\alpha}, s d^{-\alpha}\hat{h}\right)\right] \label{Eq_Eta_a} \\
		&= \lambda' \pi d^2 - \lambda' \pi d^2 \mathbb{E}_{\hat{h}}\left[\pFq{1}{1}{-\frac{2}{\alpha}}{1-\frac{2}{\alpha}}{-s d^{-\alpha} \hat{h}}\right]\label{Eq_Eta_b}\\
		&= \lambda' \pi d^2 - \lambda' \pi d^2 \pFq{2}{1}{1,-\frac{2}{\alpha}}{1-\frac{2}{\alpha}}{-s \mu d^{-\alpha}}, \label{Eq_Eta_c}
\end{align}
where \cite[Eq.~(8.351)]{Gradshteyn00} is exploited to reach \eqref{Eq_Eta_b}, and \eqref{Eq_Eta_c} follows the fact $\hat{h} \sim \exp(\mu)$.

By using a similar method to that in \cite{7412737, 8490204}, the recursive relations between the derivatives of $L_{I_1}(s)$ can be attained, based on which a compact Toeplitz matrix expression for the coverage probability is finally derived. Specifically, let $q_n\triangleq \frac{(-s)^n}{n!}L_{I_1}^n(s)$. Then, it is clear that
\begin{align} \label{Eq_q0_t0}
	q_0 &\triangleq L_{I_1}(s)|_{s=\frac{1}{\Omega} m \gamma d^{\alpha}}\nonumber\\
		&= \exp\left\{\lambda' \pi d^2 - \lambda' \pi d^2 \pFq{2}{1}{1, -\frac{2}{\alpha}}{1-\frac{2}{\alpha}}{-\frac{1}{\Omega} m \mu \gamma} \right\}\nonumber\\
		&
		= \exp(t_0),
\end{align}
where $t_0 \triangleq \lambda' \pi d^2 - \lambda' \pi d^2 \pFq{2}{1}{1, -\frac{2}{\alpha}}{1-\frac{2}{\alpha}}{-\frac{1}{\Omega} m \mu \gamma}$. Next, combining \eqref {Eq_Eta_a0} with \eqref{Eq_q0_t0} yields $L_{I_1}^{(1)}(s) = \eta^{(1)}(s)L_{I_1}(s)$. Afterwards, by recursion, for all $n \geq 1$, we have
\begin{equation}
	L_{I_1}^{(n)}(s) = \frac{\mathrm{d}^{n-1}}{\mathrm{d}s^{n-1}}L_{I_1}^{(1)}(s)
			        = \sum_{i=0}^{n-1}{n-1 \choose i}\eta^{n-i}(s)L_{I_1}^{(i)}(s),
\end{equation}
followed by
\begin{equation} \label{Eq-qn}
	\frac{(-s)^n}{n!}L_{I_1}^{(n)}(s) = \sum_{i=0}^{n-1}\frac{n-i}{n}\frac{(-s)^{n-i}}{(n-i)!}\eta^{(n-i)}(s) \frac{(-s)^{i}}{i!}L_{I_1}^{(i)}(s).
\end{equation}
Let $q_n = \frac{(-s)^n}{n!}L_{I_1}^{(n)}(s)$. Then, for all $n \geq 1$, \eqref{Eq-qn} implies that
\begin{equation} \label{Eq_Recursive}
	q_n = \sum_{i=0}^{n-1}\frac{n-i}{n}t_{n-i}\,q_i,
\end{equation}
wher
\begin{align} \label{Eq_tk}
	t_n &= \frac{(-s)^n}{n!}\eta^{(n)}(s)\bigg|_{s=\frac{1}{\Omega}m \gamma d^{\alpha}} = \frac{(-s)^n}{n!} \bigg( \lambda' \pi d^2  \nonumber\\
	&\quad \left. {}- \lambda' \pi d^2 \pFq{2}{1}{1, -\frac{2}{\alpha}}{1-\frac{2}{\alpha}}{-s \mu d^{-\alpha}} \right)^{(n)}\bigg|_{s=\frac{1}{\Omega}m \gamma d^{\alpha}} \nonumber\\
		&= {}-\lambda' \pi d^2\frac{2}{(2-n\alpha)} \left(\frac{1}{\Omega} m \mu \right)^n \gamma^n \nonumber\\
		&\quad \ \times \pFq{2}{1}{n+1, n-\frac{2}{\alpha}}{n+1-\frac{2}{\alpha}}{-\frac{1}{\Omega} m \mu \gamma}.
\end{align}
Combining \eqref{Eq_q0_t0} with \eqref{Eq_tk} yields the intended \eqref{Eq_q_n}.
	
Next, to explicitly express $q_n$, we define two power series as follows:
\begin{equation} \label{Eq_Tz-0}
	T(z)  \triangleq \sum_{n=0}^{\infty}t_n z^n, \quad Q(z) \triangleq \sum_{n=0}^{\infty}q_n z^n.
\end{equation}
By taking the first-order derivative of $T(z)$ and $Q(z)$, we have
\begin{equation} \label{Eq_Tz}
	T^{(1)}(z)=\sum_{n=0}^{\infty}(n+1)t_{n+1} z^{n}, \quad Q^{(1)}(z)=\sum_{n=0}^{\infty}n q_n\,z^{n-1}.
\end{equation}
Combining \eqref{Eq_Recursive}, \eqref{Eq_Tz-0} and \eqref{Eq_Tz} yields
\begin{align} \label{Eq_differentialEquation}
	T^{(1)}(z) Q(z) &= \sum_{n=0}^{\infty}\sum_{i=0}^{n-1}(n-i)t_{n-i}z^{n-1}q_i z^{i-1} \nonumber\\
	&=\sum_{n=0}^{\infty}n q_n\,z^{z-1}=Q^{(1)}(z),
\end{align}
which implies that
\begin{equation}
	Q(z) = a\exp(T(z)).
\end{equation}
By recalling \eqref{Eq_q0_t0}, $q_0 = \exp\left(t_0\right)$ leads to $a=1$ and, consequently, the coverage probability given by \eqref{Eq-CP-CaseI} can be explicitly computed as
\begin{align}
	\mathcal{P}_1(\gamma, \lambda, \alpha)
		&= \mathbb{E}_{d} \left[ \sum_{n=0}^{m-1}{q}_n \right]
		= \mathbb{E}_d \left[ \sum_{n=0}^{m-1}\frac{1}{n!} {Q}^{(n)}(z)\bigg|_{z=0} \right] \nonumber\\
		&= \mathbb{E}_d\left[ \sum_{n=0}^{m-1} \frac{1}{n!} \frac{\mathrm{d}^n}{\mathrm{d}z^n}\exp(T(z))\bigg|_{z=0} \right].
\end{align}
Finally, by using a similar technique to that in \cite{8490204}, the first $m-1$ coefficients of the power series $\exp(Q(z))$ can be derived and expressed as the first column of the matrix exponential $\exp\left(\bm{Q}\right)$, which completes the proof.
\section{Proof of Corollary~\ref{Corollary_TypeI}}
\label{Appendix-A}
With a single antenna at each BS, i.e., $M = 1$, it is clear that $\|h_i\|$ is Rayleigh distributed and, as per \eqref{Eq_PDF_T1-Approx-2b}, we get $m =3$. Then, the coverage probability given by \eqref{Eq-Coverage1} reduces to
\begin{align} \label{Eq-Ray_Coverage}
	\mathcal{P}_{1}(\gamma,\lambda, \alpha)
	 &= \int_{x > 0}  f_d(x) \left[L_{I_1}(s) - \frac{m \gamma d^\alpha}{\Omega} \frac{\partial L_{I_1}(s)}{\partial s} \right. \nonumber\\
	 &\quad \left. {}+ \frac{\left(m \gamma d^\alpha\right)^2}{2\Omega^2} \frac{\partial^2 L_{I_1}(s)}{\partial s^2} \bigg|_{s=\frac{1}{\Omega}m \gamma d^\alpha}  \right] {\rm d}x.
\end{align}
Next, we calculate the three terms in the square brackets on the right-hand side of \eqref{Eq-Ray_Coverage}. To start with, according to \eqref{Laplace_I_r}, it is straightforward that
\begin{align} \label{Eq-20}
	\lefteqn{L_{I_1}{\left(\frac{1}{\Omega}m \gamma d^\alpha\right)}} \nonumber \\
		& =  \exp\left(\frac{2 \mu \lambda' \pi  m\gamma d^2}{(2-\alpha)\Omega} \pFq{2}{1}{1, 1-\frac{2}{\alpha}}{2-\frac{2}{\alpha}}{-\frac{\mu m\gamma }{\Omega}} \right) \nonumber  \\
		& =  \exp \left( -\lambda' \pi d^2 V_1 \right),
\end{align}
where $V_1$ is shown in \eqref{Eq-V1}. Then, by recalling the first-order derivative of Gaussian hypergeometric function \cite[Eq. (7.2.1.10)]{Prudnikov86}, we have
\begin{align}
	\frac{\partial L_{I_1}(s)}{\partial s}
		&=  L_{I_1}(s) \left\{\frac{2\mu \lambda' \pi  d^{2-\alpha}}{2- \alpha} \pFq{2}{1}{1, 1-\frac{2}{\alpha}}{2-\frac{2}{\alpha}}{-\mu d^{-\alpha}  s} \right. \nonumber\\
		&\quad \left. {}+  \frac{\mu^2 \lambda' \pi  d^{2-2\alpha}s}{\alpha -1} \pFq{2}{1}{2, 2-\frac{2}{\alpha}}{3-\frac{2}{\alpha}}{- \mu d^{-\alpha}  s} \right\}.   \label{Eq-21}
\end{align}
Substituting $s = \frac{1}{\Omega}m \gamma d^\alpha$ into \eqref{Eq-21} yields
\begin{align}
	 \lefteqn{\frac{\partial L_{I_1}(s)}{\partial s}\bigg |_{s = \frac{1}{\Omega}m \gamma d^\alpha}} \nonumber \\
		 &=  L_{I_1}\left(\frac{1}{\Omega}m \gamma d^\alpha\right) \left[\frac{2\mu \lambda' \pi  d^{2-\alpha}}{2- \alpha} \pFq{2}{1}{1, 1-\frac{2}{\alpha}}{2-\frac{2}{\alpha}}{-\frac{\mu m\gamma }{\Omega}} \right. \nonumber\\
		&\quad \left. {}+  \frac{\mu^2 \lambda' \pi ^2 d^{2-\alpha} m \gamma}{(\alpha -1)\Omega} \pFq{2}{1}{2, 2-\frac{2}{\alpha}}{3-\frac{2}{\alpha}}{-\frac{\mu m\gamma }{\Omega}} \right]\nonumber\\
		& =  \lambda' \pi d^{2-\alpha}V_2 L_{I_1}\left(\frac{1}{\Omega}m \gamma d^\alpha\right),  \label{Eq-23}
\end{align}
where $V_2$ is shown in \eqref{Eq-V2}. By using a similar approach as above, we have
\begin{align} \label{Eq-25}
	\lefteqn{\frac{\partial^2 L_{I_1}(s)}{\partial s^2} \bigg|_{s=\frac{1}{\Omega}m \gamma d^\alpha}} \nonumber \\
		& = \left((\lambda' \pi)^2 d^{4-2\alpha}V_2^2  + \lambda' \pi d^{2-2\alpha}V_3 \right) L_{I_1}\left(\frac{1}{\Omega}m \gamma d^\alpha\right),
\end{align}
where $V_3$ is expressed as \eqref{Eq-V3}. Finally, substituting Eqs.~\eqref{Eq_PDF_d}, \eqref{Eq-20}, \eqref{Eq-23}, and \eqref{Eq-25} into \eqref{Eq-Ray_Coverage}, and performing some calculus, yields the intended \eqref{Eq-CoMP-Rayleigh}.
\section{Proof of Theorem~\ref{Theorem_TypeI_2}}
\label{Appendix-B}
Let $G \triangleq \max\{g_1', g_2', g_3'\}$. Since $g_i' \triangleq \|\bm{h}_i\|^2$, $i = 1, 2, 3$, are i.i.d. Gamma random variables, with parent PDF and CDF given by
\begin{equation}
    f_{g_i'}(x)= \frac{1}{\Gamma(M)}x^{M-1}\exp(-x),
\end{equation}
and
\begin{equation} \label{Eq-CDF_G}
    F_{g_i'}(x) = 1-\frac{\Gamma(M, x)}{\Gamma(M)} = 1-Q(M, x),
\end{equation}
respectively, then by recalling the theory of order statistics, the CCDF of $G$ is expressed as
\begin{equation} \label{Eq-CCDF_G}
    	{\rm{Pr}}(G > x) = 1- \left( F_{g_i'}(x) \right)^{3}.
\end{equation}
Consequently, the coverage probability can be computed as
\begin{align}
    {\rm{Pr}}(\gamma, \lambda, \alpha)
    	& =  \int_{x > 0} {\rm{Pr}} \left[\frac{r^{-\alpha} G}{I_2} > \gamma | d \right] f_d(x) \, {\rm d} x \nonumber \\
    	& =  \int_{x > 0} {\rm{Pr}} \left[ G > \gamma d^{\alpha} I_2 | d \right] f_d(x) \, {\rm d}x. \label{Coverage_Max_OneBase}
\end{align}
By virtue of \eqref{Eq-CDF_G} and \eqref{Eq-CCDF_G}, the conditional CCDF of $G$ needed in \eqref{Coverage_Max_OneBase} can be expressed as
\begin{align} \label{Max_G}
  & {\rm{Pr}} \left[G > \gamma d^{\alpha} I_2 | d \right]
     =  \mathbb{E}_{I_2} \left[ 3Q\left(M, \gamma d^{\alpha} I_2 | d\right) \right. \nonumber \\
 & \qquad \left. - 3Q^2\left(M, \gamma d^{\alpha}I_2 |d \right) + Q^3\left(M, \gamma d^{\alpha} I_2 |d\right) \right],
\end{align}
where
\begin{equation}
    	\mathbb{E}_{I_2} \left[Q\left(M, \gamma d^{\alpha} I_2 |d \right) \right]
		= \sum_{k=0}^{M - 1}\frac{1}{k!}\left(-\gamma d^{\alpha} \right)^k \frac{\partial^k L_{I_2}(s)}{\partial s^k} \bigg|_{s = \gamma d^{\alpha}}.
\end{equation}

By using a similar approach to the proof of Theorem~\ref{Theorem_TypeI}, the conditional expectation of $Q\left(M, \gamma d^{\alpha} I_2 \right)$ with respect to $I_2$ given $d$ can be explicitly expressed as	
\begin{eqnarray}\label{Eq_Ls_I2_Toeplitz}
\mathbb{E}_{I_2} \left[Q\left(M, \gamma d^{\alpha} I_2 |d\right)\right]
= \left \| \exp(\bm{Q}'(d))  \right \|_1,
\end{eqnarray}
where $\bm{Q}'$ is an $M \times M$ lower triangular Toeplitz matrix with none-zero entries shown in \eqref{Eq_q_n_2}. Finally, substituting \eqref{Max_G} into \eqref{Coverage_Max_OneBase} yields the desired \eqref{Coverage_CoMp_OneBase}.

\bibliographystyle{IEEEtran}
\bibliography{References}

\begin{IEEEbiography}
	[{\includegraphics[width=1in, height=1.25in, clip, keepaspectratio]{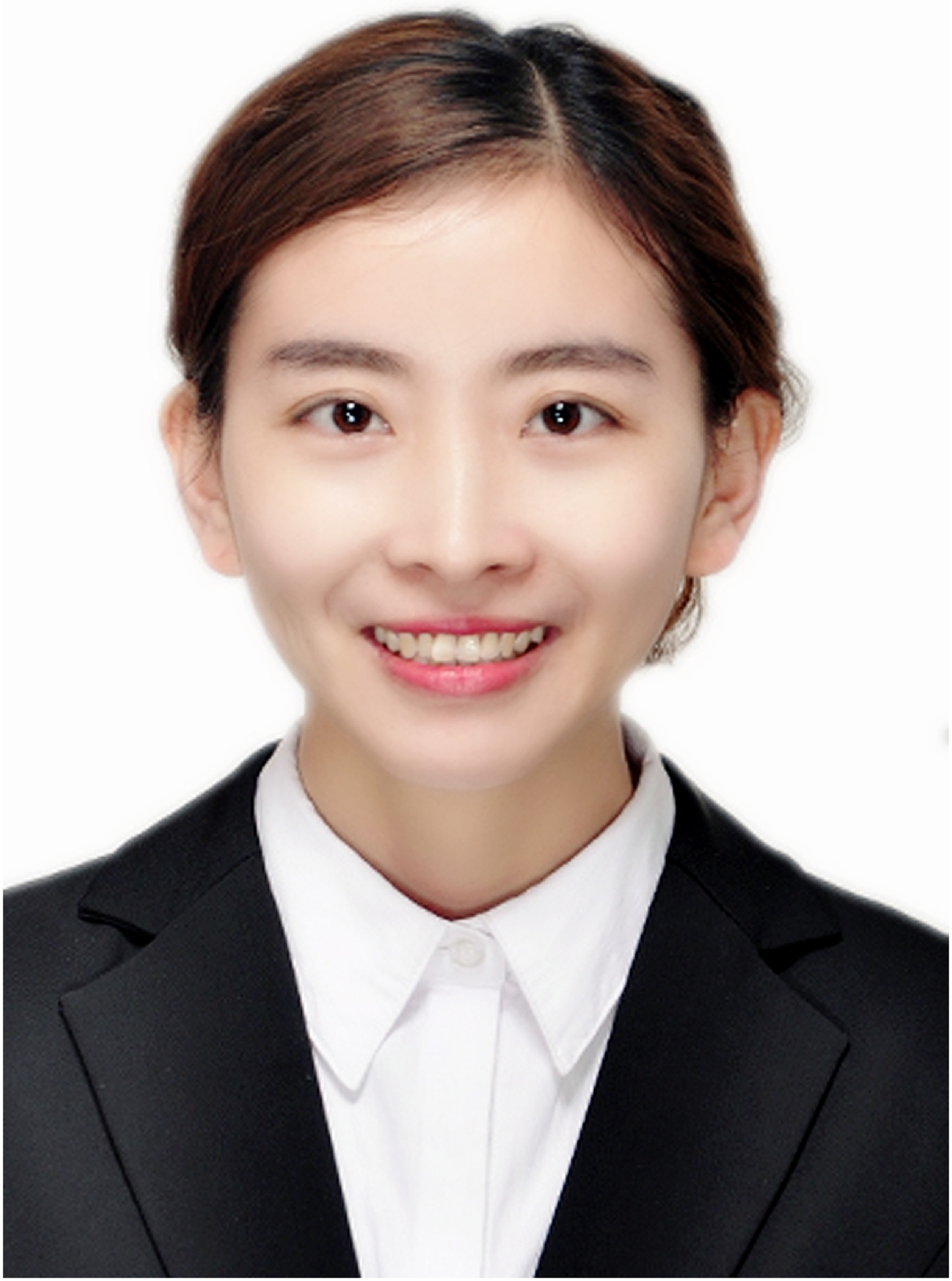}}]{Yan Li}  received the B.S. degree in Electronic Information Engineering from Hunan Normal University, Changsha, China, in 2013, and the M.S. degree in Electronics and Communication Engineering from Sun Yat-sen University, Guangzhou, China, in 2016. She is currently working towards the Ph.D. degree in Information and Communication Engineering at Sun Yat-sen University. Her research interests include modeling and analysis of cellular networks based on stochastic geometry theory, as well as cooperative communications.
\end{IEEEbiography}

\begin{IEEEbiography}
 [{\includegraphics[width=1in, height=1.25in, clip, keepaspectratio]{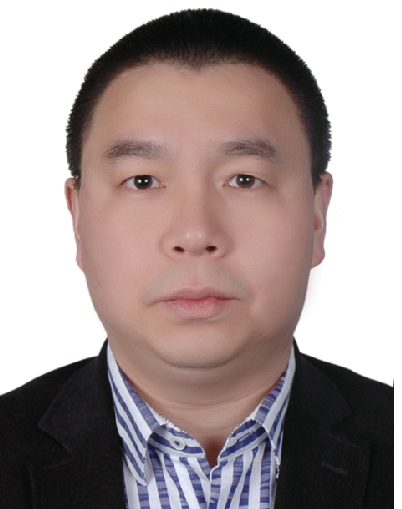}}]{Minghua Xia} (M'12) received his Ph.D. degree in Telecommunications and Information Systems from Sun Yat-sen University, Guangzhou, China, in 2007. Since 2015, he has been a Professor with Sun Yat-sen University.

From 2007 to 2009, he was with the Electronics and Telecommunications Research Institute (ETRI) of South Korea, Beijing R\&D Center, Beijing, China, where he worked as a member and then as a senior member of engineering staff. From 2010 to 2014, he was in sequence with The University of Hong Kong, Hong Kong, China; King Abdullah University of Science and Technology, Jeddah, Saudi Arabia; and the Institut National de la Recherche Scientifique (INRS), University of Quebec, Montreal, Canada, as a Postdoctoral Fellow. His research interests are in the general areas of wireless communications and signal processing.

Dr. Xia received the Professional Award at the IEEE TENCON, held in Macau, in 2015. He was recognized as Exemplary Reviewer by {\scshape IEEE Transactions on Communications} in 2014, {\scshape IEEE Communications Letters} in 2014, and {\scshape IEEE Wireless Communications Letters} in 2014 and 2015. Dr. Xia serverd as TPC Symposium Chair of IEEE ICC'2019 and now serves as Associate Editor for the {\scshape IEEE Transactions on Cognitive Communications and Networking} and the {\scshape IET Smart Cities}.
\end{IEEEbiography}

\begin{IEEEbiography}
[{\includegraphics[width=1in, height=1.25in, clip, keepaspectratio]{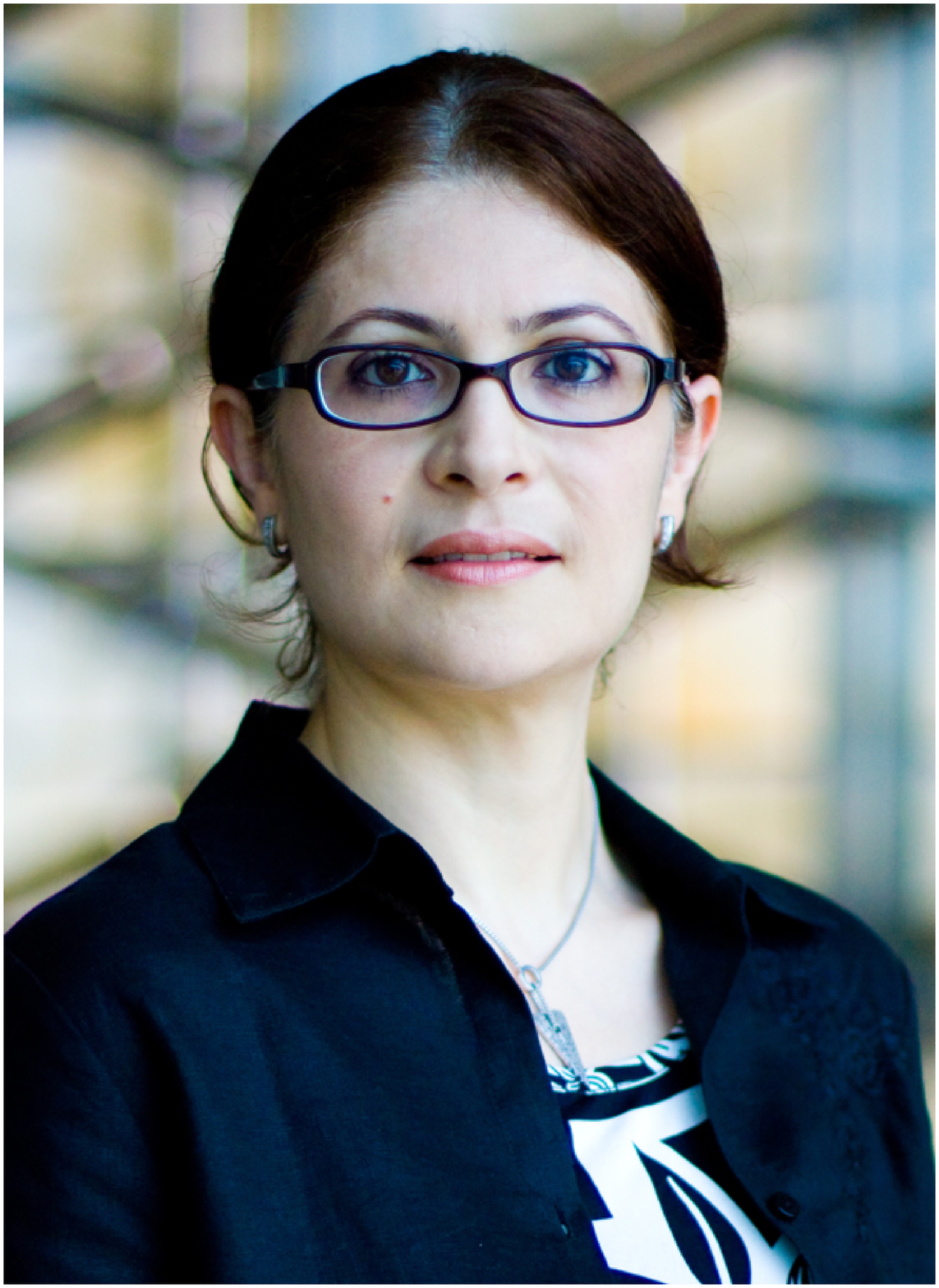}}]
{Sonia A\"{\i}ssa} (S'93-M'00-SM'03-F'19) received her Ph.D. degree in Electrical and Computer Engineering from McGill University, Montreal, QC, Canada, in 1998. Since then, she has been with the Institut National de la Recherche Scientifique-{\it Energy, Materials and Telecommunications} Center (INRS-EMT), University of Quebec, Montreal, QC, Canada, where she is a Full Professor.

From 1996 to 1997, she was a Researcher with the Department of Electronics and Communications of Kyoto University, and with the Wireless Systems Laboratories of NTT, Japan. From 1998 to 2000, she was a Research Associate at INRS-EMT. In 2000-2002, while she was an Assistant Professor, she was a Principal Investigator in the major program of personal and mobile communications of the Canadian Institute for Telecommunications Research, leading research in radio resource management for wireless networks. From 2004 to 2007, she was an Adjunct Professor with Concordia University, Canada. She was Visiting Invited Professor at Kyoto University, Japan, in 2006, and at Universiti Sains Malaysia, in 2015. Her research interests include the modeling, design and performance analysis of wireless communication systems and networks.

Dr. A\"{\i}ssa is the Founding Chair of the IEEE Women in Engineering Affinity Group in Montreal, 2004-2007; acted as TPC Symposium Chair or Cochair at IEEE ICC '06 '09 '11 '12; Program Cochair at IEEE WCNC 2007; TPC Cochair of IEEE VTC-spring 2013; TPC Symposia Chair of IEEE Globecom 2014; TPC Vice-Chair of IEEE Globecom 2018; and serves as the TPC Chair of IEEE ICC 2021. Her main editorial activities include: Editor, {\scshape IEEE Transactions on Wireless Communications}, 2004-2012; Associate Editor and Technical Editor, {\scshape IEEE Communications Magazine}, 2004-2015; Technical Editor, {\scshape IEEE Wireless Communications Magazine}, 2006-2010; and Associate Editor, {\it Wiley Security and Communication Networks Journal}, 2007-2012. She currently serves as Area Editor for the {\scshape IEEE Transactions on Wireless Communications}. Awards to her credit include the NSERC University Faculty Award in 1999; the Quebec Government FRQNT Strategic Faculty Fellowship in 2001-2006; the INRS-EMT Performance Award multiple times since 2004, for outstanding achievements in research, teaching and service; and the Technical Community Service Award from the FRQNT Centre for Advanced Systems and Technologies in Communications, 2007. She is co-recipient of five IEEE Best Paper Awards and of the 2012 IEICE Best Paper Award; and recipient of NSERC Discovery Accelerator Supplement Award. She served as Distinguished Lecturer of the IEEE Communications Society and Member of its Board of Governors in 2013-2016 and 2014-2016, respectively. Professor A\"{\i}ssa is a Fellow of the Canadian Academy of Engineering.
\end{IEEEbiography}

\vfill

\end{document}